\documentclass[a4paper,11pt]{article}
\usepackage{jheppub} 
\usepackage[T1]{fontenc}
\usepackage{booktabs}
\usepackage[dvips,table]{xcolor} 
\usepackage{xspace}
\usepackage{dcolumn}
\usepackage{hyperref}
\usepackage{caption}
\usepackage
[subrefformat=parens,position=top,skip=-15pt,margin=15pt,justification=justified,singlelinecheck=false]
{subcaption}

\newcommand{\eq}[1]{\begin{equation} #1 \end{equation}}

\allowdisplaybreaks[1]

\newcommand{\Pl}{\ell}

\def\reffi#1{\mbox{Figure~\ref{#1}}}
\def\reffis#1{\mbox{Figures~\ref{#1}}}
\def\refta#1{\mbox{Table~\ref{#1}}}

\def\refse#1{\mbox{Section~\ref{#1}}}
\def\refses#1{\mbox{Sections~\ref{#1}}}

\def\citere#1{\mbox{Ref.~\cite{#1}}}
\def\citeres#1{\mbox{Refs.~\cite{#1}}}

\newcommand{\ri}{\mathrm i}

\def\be{\begin{equation}}
\def\ee{\end{equation}}

\newcommand{\PH}{\ensuremath{\text{H}}\xspace}
\newcommand{\Pj}{\ensuremath{\text{j}}\xspace}
\newcommand{\Pp}{\ensuremath{\text{p}}\xspace}
\newcommand{\Pe}{\ensuremath{\text{e}}\xspace}
\newcommand{\Pb}{\ensuremath{\text{b}}\xspace}

\newcommand{\Pt}{\ensuremath{\text{t}}\xspace}
\newcommand{\Pu}{\ensuremath{\text{u}}\xspace}
\newcommand{\Pd}{\ensuremath{\text{d}}\xspace}
\newcommand{\Ps}{\ensuremath{\text{s}}\xspace}
\newcommand{\Pc}{\ensuremath{\text{c}}\xspace}
\newcommand{\Pg}{\ensuremath{\text{g}}\xspace}

\newcommand{\PW}{\ensuremath{\text{W}}\xspace}
\newcommand{\PZ}{\ensuremath{\text{Z}}\xspace}

\newcommand{\Mt}{\ensuremath{m_\Pt}\xspace}

\newcommand{\MWOS}{\ensuremath{M_\PW^\text{OS}}\xspace}
\newcommand{\MW}{\ensuremath{M_\PW}\xspace}
\newcommand{\MZOS}{\ensuremath{M_\PZ^\text{OS}}\xspace}
\newcommand{\MZ}{\ensuremath{M_\PZ}\xspace}

\newcommand{\Gt}{\ensuremath{\Gamma_\Pt}\xspace}

\newcommand{\GZOS}{\ensuremath{\Gamma_\PZ^\text{OS}}\xspace}

\newcommand{\GWOS}{\ensuremath{\Gamma_\PW^\text{OS}}\xspace}

\newcommand{\GeV}{\ensuremath{\,\text{GeV}}\xspace}
\newcommand{\TeV}{\ensuremath{\,\text{TeV}}\xspace}

\newcommand{\alphas}{\ensuremath{\alpha_\text{s}}\xspace}
\newcommand{\order}[1]{\ensuremath{\mathcal{O}{\left(#1\right)}}\xspace}

\newcommand{\abs}[1]{\left|#1\right|}

\newcommand{\GF}{\ensuremath{G_\mu}}

\newcommand{\pt}{\ensuremath{p_\text{T}}\xspace}
\newcommand{\ptsub}[1]{\ensuremath{p_{\text{T},#1}}\xspace}

\renewcommand{\Re}{\mathop{\mathrm{Re}}\nolimits}

\newcommand{\fullProcess  }{\ensuremath{\Pp\Pp\to\Pe^+\nu_\Pe \mu^-\bar{\nu}_\mu\Pb\bar{\Pb}\PH}\xspace}

\newcommand{\recola}{{\sc Recola}\xspace}
\newcommand{\mocanlo}{{\sc MoCaNLO}\xspace}
\newcommand{\collier}{{\sc Collier}\xspace}
\newcommand{\madgraph}{{\sc\small MadGraph5\_aMC@NLO}\xspace}

\newcolumntype{.}{D{.}{.}{-1}}
\newcolumntype{d}[1]{D{.}{.}{#1}}

\renewcommand{\vec}[1]{\mathbf{#1}}
\colorlet{tableoverheadcolor}{gray!37.5}
\colorlet{tableheadcolor}{gray!25}
\colorlet{tablerowcolor}{gray!12.5}

\newlength{\width}
\newlength{\height}

\def\draftdate{\relax}
\def\mda{\relax}
\def\mua{\relax}
\def\mla{\relax}
\def\draft{
\def\thtystars{******************************}
\def\sixtystars{\thtystars\thtystars}
\typeout{}
\typeout{\sixtystars**}
\typeout{* Draft mode!
         For final version remove \protect\draft\space in source file *}
\typeout{\sixtystars**}
\typeout{}
\def\draftdate{\today}
\def\mua{\marginpar[\boldmath\hfil$\uparrow$]%
                   {\boldmath$\uparrow$\hfil}\color{black}%
                    \typeout{marginpar: $\uparrow$}\ignorespaces}
\def\mda{\color{red}\marginpar[\boldmath\hfil$\downarrow$]%
                   {\boldmath$\downarrow$\hfil}%
                    \typeout{marginpar: $\downarrow$}\ignorespaces}
\def\mla{\marginpar[\boldmath\hfil$\rightarrow$]%
                   {\boldmath$\leftarrow $\hfil}%
                    \typeout{marginpar: $\leftrightarrow$}\ignorespaces}
\def\Mua{\marginpar[\boldmath\hfil$\Uparrow$]%
                   {\boldmath$\Uparrow$\hfil}\color{black}%
                    \typeout{marginpar: $\uparrow$}\ignorespaces}
\def\Mda{\color{red}\marginpar[\boldmath\hfil$\Downarrow$]%
                   {\boldmath$\Downarrow$\hfil}%
                    \typeout{marginpar: $\downarrow$}\ignorespaces}
\def\Mla{\marginpar[\boldmath\hfil\textcolor{red}{$\Rightarrow$}]%
                   {\boldmath\textcolor{red}{$\Leftarrow $}\hfil}%
                    \typeout{marginpar: $\leftrightarrow$}\ignorespaces}
\overfullrule 5pt
\oddsidemargin 15mm
\marginparwidth 29mm
}

\newcommand{\change}[1]{{#1}}

\title{Higgs production in association with off-shell top--antitop pairs at NLO EW and QCD at the LHC}
\subheader{\today}

\author{Ansgar Denner$^1$,}
\author{Jean-Nicolas Lang$^1$,}
\author{Mathieu Pellen$^1$,}
\author{Sandro Uccirati$^2$}

\affiliation{$^1$ %
        Universit\"at W\"urzburg, %
        Institut f\"ur Theoretische Physik und Astrophysik, \\  %
        97074 W\"urzburg, %
        Germany%
}

\affiliation{$^2$
Universit{$\grave{a}$} di Torino e INFN, 10125 Torino, Italy
}

\emailAdd{ansgar.denner@physik.uni-wuerzburg.de}
\emailAdd{jean-nicolas.lang@physik.uni-wuerzburg.de}
\emailAdd{mathieu.pellen@physik.uni-wuerzburg.de}
\emailAdd{uccirati@to.infn.it}

\abstract{ We present NLO electroweak corrections to Higgs production
  in association with off-shell top--antitop quark pairs.  The full
  process $\Pp\Pp\to\Pe^+\nu_\Pe \mu^-\bar{\nu}_\mu\Pb\bar{\Pb} \PH$
  is considered, and hence all interference, off-shell, and
  non-resonant contributions are taken into account.  The electroweak
  corrections turn out to be below one per cent for the integrated
  cross section but can exceed $10\%$ in certain phase-space regions.
  In addition to its phenomenological relevance, the computation
  constitutes a major technical achievement as the full NLO virtual
  corrections involving up to 9-point functions have been computed
  exactly.  The results of the full computation are supported by two
  calculations in the double-pole approximation.  These also allow to
  infer the effect of off-shell contributions and emphasise their
  importance especially for the run~II of the LHC.  Finally, we
  present combined predictions featuring both NLO electroweak and QCD
  corrections in a common set-up that will help the experimental
  collaborations in their quest of precisely measuring the
  aforementioned process.  }

\begin{document}

\maketitle
\flushbottom

\newpage

\section{Introduction}
\label{sec:introduction}

Since the discovery of the Higgs boson
\cite{Aad:2012tfa,Chatrchyan:2012xdj} at the Large Hadron Collider
(LHC), significant experimental efforts have been devoted to probing
its properties.  Among these, the Higgs-boson couplings to other
particles and in particular to the top quark are of prime
importance.  The measurement of Higgs-boson production in association
with a pair of top quarks is a key input in that respect.  The
experimental measurement of this process is particularly challenging
due to the large fraction of top quarks produced by other processes.
Thus, so far only evidence for such a process has been achieved
\cite{Khachatryan:2016vau,Khachatryan:2014qaa,Aad:2015iha,Aad:2015gra,Aad:2014lma}.
This measurement allows to examine possible new-physics contributions
in the top-quark--Higgs Yukawa coupling.  Hence, state-of-the-art
predictions at next-to-leading-order (NLO) electroweak (EW) and QCD
will soon be very valuable for the experimental collaborations in
order to precisely measure the process and possibly discover
new-physics mechanisms.

For the production of a Higgs boson in association with on-shell top
quarks, already several NLO QCD computations have been performed
\cite{Beenakker:2001rj,Beenakker:2002nc,Reina:2001sf,Dawson:2003zu},
and their matching to parton showers
\cite{Frederix:2011zi,Garzelli:2011vp,Hartanto:2015uka} is also
available.
\change{Moreover, resummation of soft-gluon-emission
contributions for $\Pt\bar{\Pt}\PH$ production has been performed
to next-to-next-to-leading-logarithmic (NNLL) accuracy
\cite{Kulesza:2015vda,Broggio:2015lya,Kulesza:2016vnq,Broggio:2016lfj}.}
On the other hand, the computation of the NLO QCD
corrections for off-shell top quarks has been realised only recently
for the first time at the LHC~\cite{Denner:2015yca} and at a linear
collider~\cite{Nejad:2016bci}.  Concerning NLO EW, only computations
in the limit of on-shell top
quarks~\cite{Frixione:2014qaa,Frixione:2015zaa,Badger:2016bpw} or in
the narrow-width approximation~\cite{Yu:2014cka} are available so far.
With the present computation we go beyond the on-shell approximation
and compute for the first time the full NLO EW corrections to
Higgs-boson production in association with off-shell top quarks,
\emph{i.e.}~the complete process $\Pp\Pp\to\Pe^+\nu_\Pe
\mu^-\bar{\nu}_\mu\Pb\bar{\Pb} \PH$.  Hence, it features all
non-resonant, interference, and off-shell effects allowing to make
realistic predictions that can be directly compared to experiments.
The present article follows a series of several NLO computations
involving off-shell top quarks
\cite{Denner:2012yc,Denner:2015yca,Denner:2016jyo}.

The presented computation is also a major technical achievement.
While the Born and real contributions are of similar complexity as
those in \citeres{Denner:2012yc,Denner:2015yca,Denner:2016jyo}, the
virtual contributions are significantly more complicated.  Indeed,
this is the first time that an NLO computation involving one-loop
amplitudes with up to 9-point functions is made public.
So far, the most complicated NLO computations have been limited to 8-point functions \cite{Bern:2013gka,Denner:2016jyo,Biedermann:2016yds}.
This progress is made possible by the use of the public codes \recola \cite{Actis:2012qn,Actis:2016mpe} and \collier \cite{Denner:2014gla,Denner:2016kdg}.

In addition to calculating the NLO EW corrections to the complete
process, two computations in a double-pole approximation (DPA) are
also provided.  In the first one, two resonant W bosons are required
while in the second, two resonant top quarks are demanded.
On the one hand, the DPAs allow to
estimate the size of non-resonant and off-shell contributions and
hence to infer the validity of on-shell computations especially for
differential distributions.  Secondly, they constitute an important
cross-check of the full computation.  This is an important point as the full virtual amplitude is
particularly involved.  Both DPAs turn out to be very good
approximations of the full computation.

Finally, in order to provide state-of-the-art predictions for
Higgs-boson production in association with off-shell top quarks, we
have recomputed the NLO QCD corrections presented in
\citere{Denner:2015yca}.  In that way we can provide 
predictions featuring both NLO EW and QCD corrections to the full
process in a consistent set-up.  We present results
for two different ways of combining NLO effects.  The first one is an
additive combination while the second one is a multiplicative
combination.

In \refse{sec:calculation} details concerning the calculation are provided.
While \refses{ssec:RealCorrections} and \ref{ssec:VirtualCorrections}
are devoted to the description of real and virtual corrections,
respectively, in \refse{sec:DoublePoleApproximation} the two DPAs are
reviewed, and in \refse{sec:Validations} the validation of the
computation is described.  In \refse{sec:results}, numerical results
are presented for a centre-of-mass energy of $\sqrt{s}=13\TeV$ at the
LHC.  In particular, in \refse{ssec:InputParameters} the event
selection and the input parameters are specified, and in
\refses{ssec:IntegratedCrossSection} and
\ref{ssec:DifferentialDistributions} the results for the integrated
cross sections and differential distributions are provided.  The
comparison between the full calculation and the two DPAs is performed
in \refse{sec:ComparisonDPA} for both the total cross section and
differential distributions.  Finally, predictions for the LHC
including both NLO EW and QCD are provided in
\refse{sec:CombinationNLO}.  \refse{sec:Conclusions} contains the
conclusion.

\section{Details of the calculation}
\label{sec:calculation}

\begin{figure}
        \newcommand{\myframebox}{\framebox}
        \renewcommand{\myframebox}{\relax}
        \setlength{\parskip}{-10pt}
        \captionsetup[subfigure]{margin=5pt}
        \begin{subfigure}{0.32\linewidth}
                \myframebox{
                        \includegraphics[width=\linewidth]{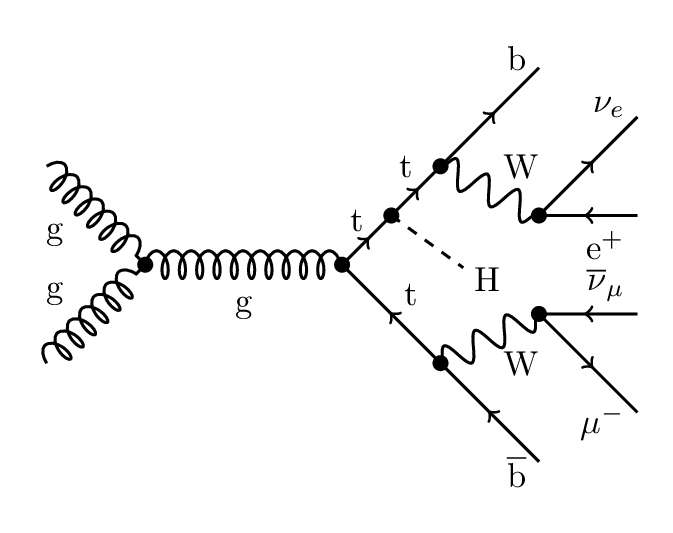}
                }
        \end{subfigure}
        \begin{subfigure}{0.37\linewidth}
                \myframebox{
                        \includegraphics[width=\linewidth]{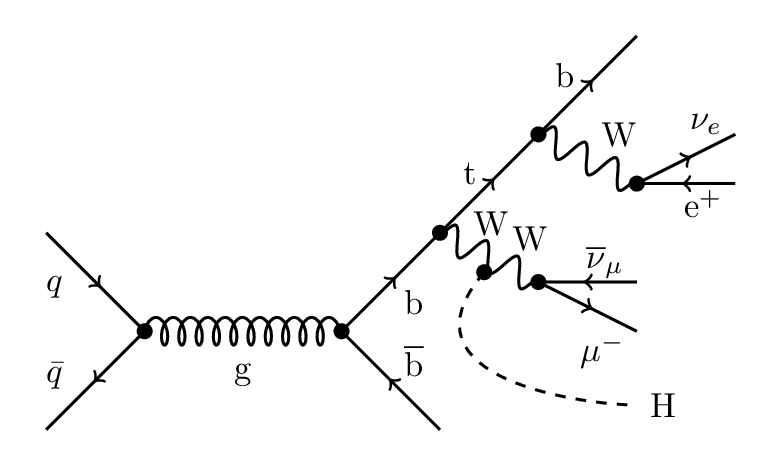}
                }
        \end{subfigure}
        \begin{subfigure}{0.27\linewidth}
                \myframebox{
                        \includegraphics[width=\linewidth]{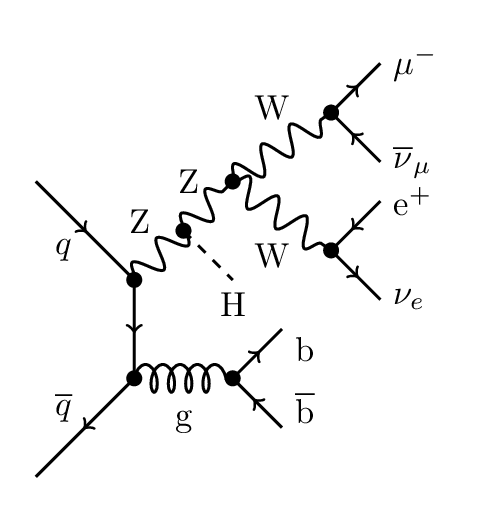}
                }
        \end{subfigure}

        \caption{\label{fig:lo_tree_feynman_diagrams}%
          Sample tree-level Feynman diagrams with two (left),
          one (middle) and no (right) top-quark resonances.  }
\end{figure}

\begin{figure}
        \newcommand{\myframebox}{\framebox}
        \renewcommand{\myframebox}{\relax}
        \setlength{\parskip}{-10pt}
        \captionsetup[subfigure]{margin=5pt}
        \center
        \begin{subfigure}{0.25\linewidth}
                \myframebox{
                        \includegraphics[width=\linewidth]{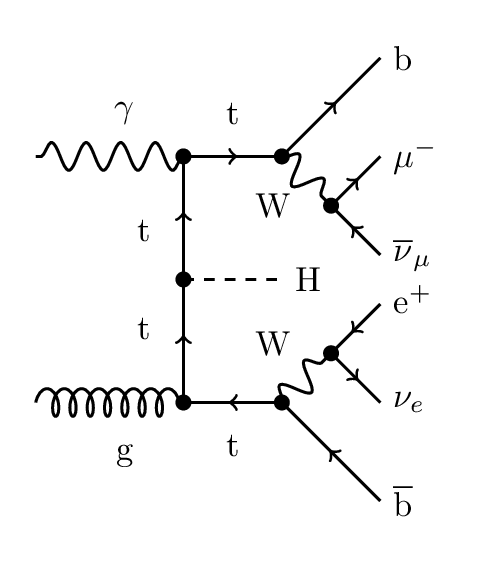}
                }
        \end{subfigure}
        \begin{subfigure}{0.25\linewidth}
                \myframebox{
                        \includegraphics[width=\linewidth]{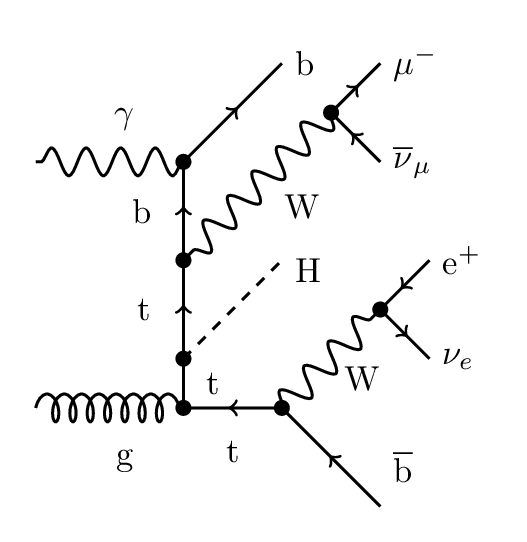}
                }
        \end{subfigure}
        \begin{subfigure}{0.42\linewidth}
                \myframebox{
                        \includegraphics[width=\linewidth]{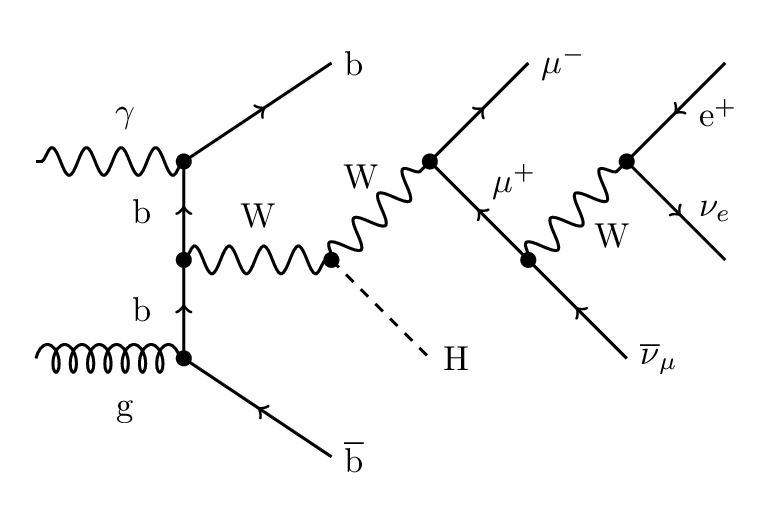}
                }
        \end{subfigure}
        \caption{\label{fig:lo_photon_feynman_diagrams}%
          Sample tree-level Feynman diagrams with photons in the initial state.  }
\end{figure}

In the present article, the NLO EW corrections to the full hadronic process
\begin{equation}\label{eqn:full_process}
        \fullProcess ,
\end{equation}
are presented. We consider the lowest-order cross section of the order
$\order{\alphas^2\alpha^{5}}$. The EW corrections to this process
consist of all possible corrections of the order
$\order{\alphas^2\alpha^{6}}$. We have neglected the LO electroweak
process of order $\order{\alpha^{7}}$, while the LO interferences of
order $\order{\alphas\alpha^{6}}$ vanish. For reference, we have
computed the Born process with initial state $\gamma {\rm g}$ of order
$\order{\alphas\alpha^{6}}$ \change{and choose to present the
  corresponding results separately.}
We have not calculated the QCD corrections to this process, which are
of order $\order{\alphas^2\alpha^{6}}$, as the photon-induced process
contributes only at the per-cent level and thus these corrections are
negligible.  The calculation presented here includes all
interferences, resonant, non-resonant, and off-shell effects of the
massive intermediate particles, \emph{i.e.}~the top quarks and the
gauge bosons.  In \reffi{fig:lo_tree_feynman_diagrams}, representative
LO diagrams featuring two, one, and no resonant top quark(s) are shown
for gluon- and quark-induced processes.  In
\reffi{fig:lo_photon_feynman_diagrams}, three diagrams of the
photon-initiated process are depicted.  Note that the quark-mixing
matrix has been assumed to be diagonal, and the bottom-quark parton
distribution function (PDF) has been neglected.

To perform the numerical integration, an in-house multi-channel Monte Carlo program {\sc MoCaNLO} \cite{MoCaNLO} has been employed.
It has already been used for several computations involving processes with high multiplicity \cite{Denner:2015yca,Denner:2016jyo,Biedermann:2016yds}.
The multi-channel phase-space integration follows the ones of \citeres{Berends:1994pv,Denner:1999gp,Dittmaier:2002ap}.
The infrared (IR) singularities in the real contributions are treated with the dipole subtraction method \cite{Catani:1996vz,Catani:2002hc,Dittmaier:1999mb}.
All matrix-elements are provided by the computer code \recola-1.1~\cite{Actis:2012qn,Actis:2016mpe}~\footnote{It is publicly available at https://recola.hepforge.org.} and the loop-integral library \collier-1.1~\cite{Denner:2014gla,Denner:2016kdg}~\footnote{It is publicly available at https://collier.hepforge.org.}.

\subsection{Real corrections}
\label{ssec:RealCorrections}

\begin{figure}
        \newcommand{\myframebox}{\framebox}
        \renewcommand{\myframebox}{\relax}
        \setlength{\parskip}{-10pt}
        \captionsetup[subfigure]{margin=5pt}
        \begin{subfigure}{0.48\linewidth}
                \myframebox{
                        \includegraphics[width=\linewidth]{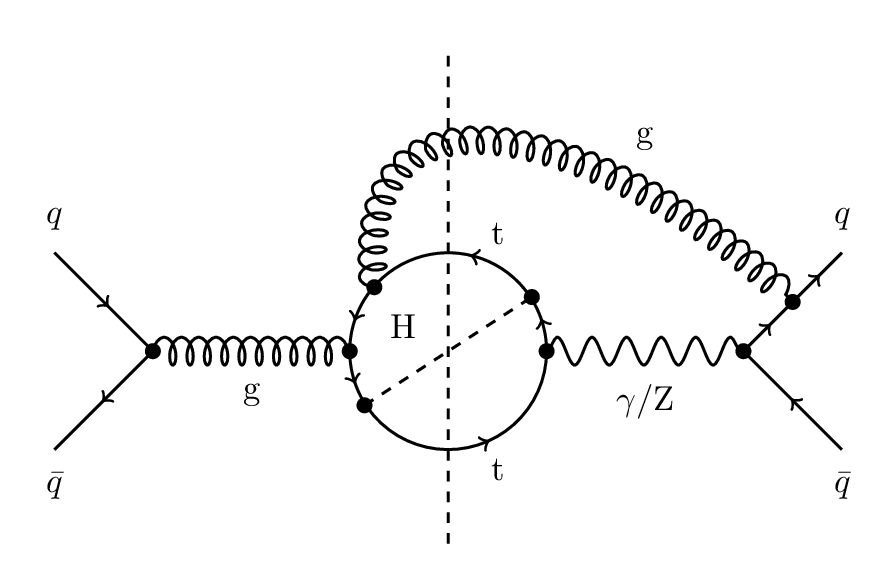}
                }
        \end{subfigure}
        \begin{subfigure}{0.48\linewidth}
                \myframebox{
                        \includegraphics[width=\linewidth]{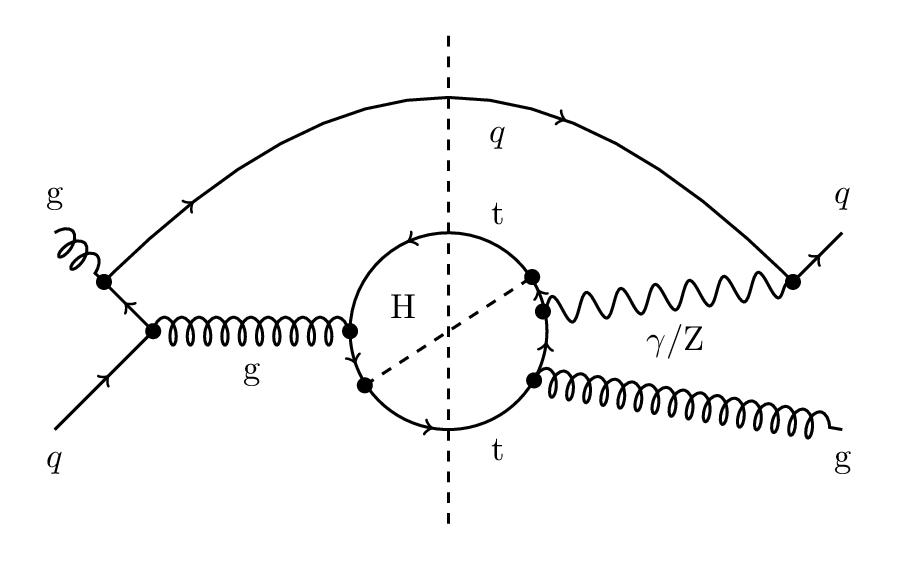}
                }
        \end{subfigure}

        \caption{\label{fig:real_int_feynman_diagrams}%
          Two Feynman diagrams squared for real corrections featuring interference between QCD and EW tree-level diagrams.
          The decay products of the top quarks are not shown as their inclusion does not alter the discussion.  }
\end{figure}

The real corrections to the process \eqref{eqn:full_process} consist of all real-radiation contributions at the order $\order{\alphas^2\alpha^{6}}$.
The first type of corrections results from photons radiated off any of the charged particles.
In addition, interferences of a QCD-mediated process emitting a gluon with its EW counterpart in the ${ q \bar{q}}$ channel must be taken into account.
This is exemplified on the left-hand side of~\reffi{fig:real_int_feynman_diagrams} where for simplicity on-shell top quarks are represented even if the computation comprises off-shell top quarks.
Finally, a last type of interference appears in the $q {\rm g}$ or $\bar{q} {\rm g}$ channel as shown on the right-hand side of~\reffi{fig:real_int_feynman_diagrams}.

The Catani--Seymour subtraction formalism for QCD \cite{Catani:1996vz,Catani:2002hc} and QED \cite{Dittmaier:1999mb} has been used for the treatment of the IR singularities.
Both the QCD and QED singularities from collinear initial-state splittings have been absorbed in the PDFs using the $\overline{\text{MS}}$ factorisation scheme.
We use the $\mathrm{LUXqed}\_\mathrm{plus}\_\mathrm{PDF4LHC15}\_\mathrm{nnlo}\_100$ set~\cite{Manohar:2016nzj} that provides a $\overline{\text{MS}}$ photon PDF.
Note that all ingredients needed for the real-subtracted part (squared amplitudes for the real-correction processes as well as the colour- and spin-correlated squared amplitudes) have been obtained from the computer code \recola \cite{Actis:2012qn,Actis:2016mpe}.

\subsection{Virtual corrections}
\label{ssec:VirtualCorrections}

\begin{figure}
        \newcommand{\myframebox}{\framebox}
        \renewcommand{\myframebox}{\relax}
        \setlength{\parskip}{-10pt}
        \captionsetup[subfigure]{margin=5pt}
        \begin{subfigure}{0.48\linewidth}
                \myframebox{
                        \includegraphics[width=\linewidth]{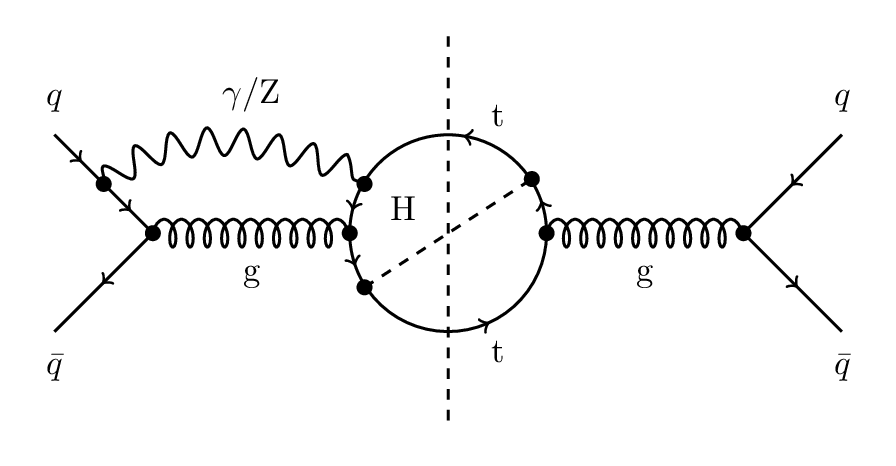}
                }
        \end{subfigure}
        \begin{subfigure}{0.48\linewidth}
                \myframebox{
                        \includegraphics[width=\linewidth]{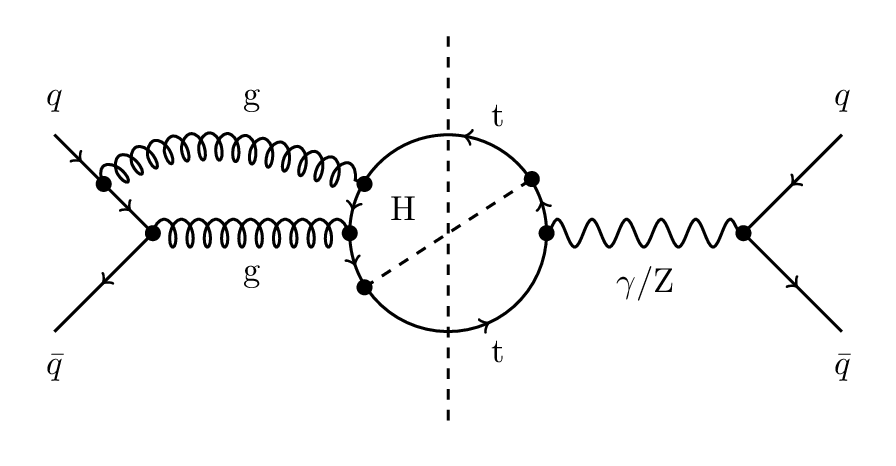}
                }
        \end{subfigure}
        \caption{\label{fig:loop_int_feynman_diagrams}%
          Sample one-loop Feynman diagrams squared.
          On the left-hand side, the diagram represents an EW correction to the QCD process which can also be seen as a QCD correction to the EW amplitude interfered with the QCD amplitude.
          On the right-hand side, a QCD correction to the QCD amplitude interfered with the EW amplitude is displayed.
          The decay products of the top quarks are not shown as their inclusion does not alter the discussion.  }
\end{figure}

\begin{figure}
        \newcommand{\myframebox}{\framebox}
        \renewcommand{\myframebox}{\relax}
        \setlength{\parskip}{-10pt}
        \captionsetup[subfigure]{margin=0pt}
        \center
        \begin{subfigure}{0.38\linewidth}
                \myframebox{
                        \includegraphics[width=\linewidth]{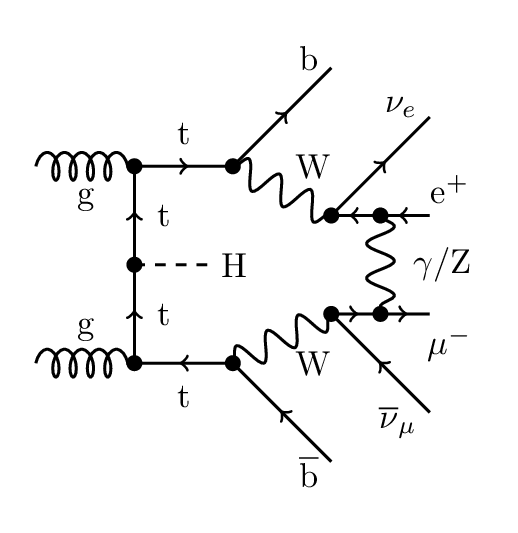}
                }
                \label{fig:virt_2tops_gg_tchannel} 
        \end{subfigure}
        \begin{subfigure}{0.51\linewidth}
                \myframebox{
                        \includegraphics[width=\linewidth]{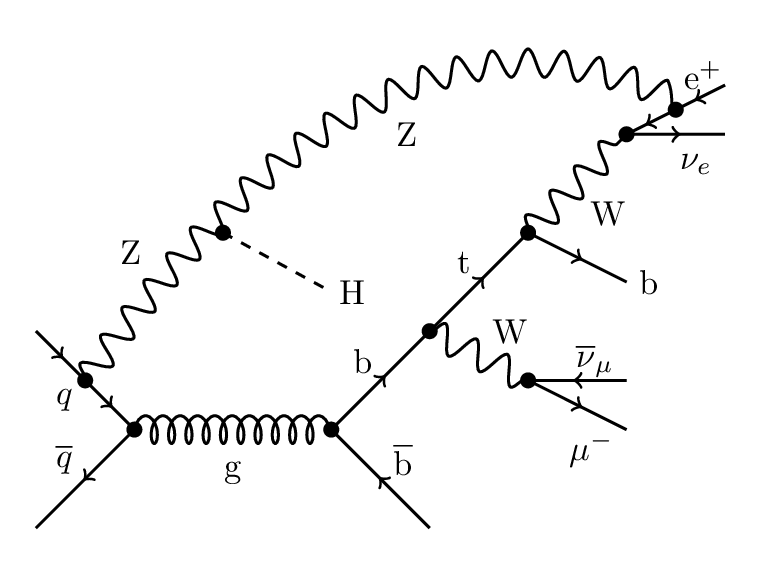}
                }
                \label{fig:virt_2tops_gg_schannel} 
        \end{subfigure}
        \caption{\label{fig:nlo_loop_feynman_diagrams}%
                Example of nonagon and octagon one-loop Feynman diagrams.
        }
\end{figure}

Analogously to the real corrections, two types of virtual corrections
must be taken into account.  The first type arises from
one-loop amplitudes of order $\order{\alphas\alpha^{7/2}}$ interfered
with tree amplitudes of order $\order{\alphas\alpha^{5/2}}$.  For the
$\bar{q}q$ channels, the second type of corrections results from the
interference of one-loop amplitudes of order
$\order{\alphas^2\alpha^{5/2}}$ with tree amplitudes of order
$\order{\alpha^{7/2}}$.  A sketch of these two types of corrections is
shown in~\reffi{fig:loop_int_feynman_diagrams}.
As for the interferences in the real corrections, the decay products
of the top quarks are not shown since their inclusion does not alter the
discussion.  Two of the most complicated loop diagrams (8- and 9-point
functions) are displayed in~\reffi{fig:nlo_loop_feynman_diagrams}.
These virtual contributions are provided by the matrix-element
generator \recola \cite{Actis:2012qn,Actis:2016mpe} in the
't~Hooft--Feynman gauge in dimensional regularisation.  
The \collier library \cite{Denner:2014gla,Denner:2016kdg} is used to
calculate the one-loop scalar
\cite{'tHooft:1978xw,Beenakker:1988jr,Dittmaier:2003bc,Denner:2010tr}
and tensor integrals
\cite{Passarino:1978jh,Denner:2002ii,Denner:2005nn} numerically.
Note that this is the first time that a NLO computation featuring 9-point functions has been made public.
As discussed in \refse{sec:Validations}, the 9-point functions yield sizeable contributions to the full process.
This demonstrates the ability of the computer codes \recola and \collier to provide fast and reliable one-loop amplitudes for complicated processes.

All resonant massive particles (top quarks, Z bosons and W bosons) are treated in the complex-mass scheme \cite{Denner:1999gp,Denner:2005fg,Denner:2006ic}, where the masses of the unstable particles as well as the weak mixing angle are complex quantities,
\begin{equation}
\label{e:WZcms}
 \overline{M}^2_\PW=\MW^2-\ri\MW\Gamma_\PW,\qquad
 \overline{M}^2_\PZ=\MZ^2-\ri\MZ\Gamma_\PZ, \qquad \text{and} \qquad
\cos\theta_{\mathrm{w}}=\frac{ \overline{M}_\PW}{ \overline{M}_\PZ}.
\end{equation}

\subsection{Double-pole approximation}
\label{sec:DoublePoleApproximation}

In \citere{Denner:2016jyo}, two DPAs have been presented for the
process $\Pp\Pp\to\Pe^+\nu_\Pe \mu^-\bar{\nu}_\mu\Pb\bar{\Pb}$ at NLO
EW.  As the Higgs boson is neither electrically charged nor charged
under QCD, all formulas presented in \citere{Denner:2016jyo} can be
applied to $\Pp\Pp\to\Pe^+\nu_\Pe \mu^-\bar{\nu}_\mu\Pb\bar{\Pb} \PH$
straightforwardly.  Therefore, we do not provide a detailed
description here, and the interested reader is referred to
\citere{Denner:2016jyo} and references therein.

The DPA serves two key purposes.  First, since
it selects resonant contributions, one can infer the effect of
non-resonant and off-shell contributions upon comparing with the full
calculation.
This, in particular, reveals if on-shell approximations can
approximate satisfactorily the full process.  Second, since the DPA
turns out to be a very good approximation to the full calculation (see
below), it serves as a check of the full virtual corrections.

As opposed to on-shell approximations, the DPA accounts for some
off-shell effects.  Indeed, the resonant propagator and the phase
space are exactly taken into account, while the rest of the matrix element is
expanded about the resonance poles.  Finally, the DPA also accounts
for resonant non-factorisable corrections.  Following
\citeres{Denner:2000bj,Accomando:2004de}, the pole approximation
has been applied only to the virtual corrections.  Thus, in the LO and
all real contributions no pole approximation is utilised.  But in
order to cancel the IR singularities originating from the virtual
corrections, one has to apply the on-shell projection to the terms
containing the $I$-operator in the integrated dipole contribution.\footnote{The $I$-, $P$-, and $K$-operator are defined in Ref.~\cite{Catani:1996vz}.}  On
the other hand, the $P$- and $K$-operator terms are still evaluated
with the off-shell kinematics like the real corrections.  This
introduces a mismatch, which is of the order of the intrinsic error of
the DPA.

The process $\Pp\Pp\to\Pe^+\nu_\Pe \mu^-\bar{\nu}_\mu\Pb\bar{\Pb} \PH$
is dominated by the production of two top quarks that decay into
bottom quarks and lepton--neutrino pairs via W bosons.  Requiring
either two resonant top quarks or two resonant W bosons is thus
expected to approximate well the full process.
We have studied two different DPAs for the process
\eqref{eqn:full_process}: In one case (WW DPA), two resonant W~bosons
are demanded, while in the second case (tt DPA), two resonant top
quarks are required.  The momenta of the resonant particles entering
the matrix elements have to be projected on shell in order to ensure
gauge invariance.  The two on-shell projections are identical to the
ones of \citere{Denner:2016jyo}.  For the WW DPA, the on-shell
projection is designed such that the invariants of the top quarks are
left untouched.  Analogously, for the on-shell projection of tt DPA,
the invariants of the W bosons are not modified.

{
In the DPA, two different kinds of corrections appear,
factorisable and non-factorisable corrections.  The factorisable
virtual corrections can be uniquely attributed either to the
production of the resonant particles or to their decays.  For a pole
approximation (PA) with $r$ resonances ($r=2$ for a DPA), they can be
written as \cite{Dittmaier:2015bfe}
}
\begin{align}
        \mathcal{M}_{\mathrm{virt,fact,PA}} & =
        \begin{aligned}[t]
                & \sum_{\lambda_1,\ldots,\lambda_r} \left( \prod^r_{i=1} \frac{1}{K_i}\right) \Bigg[ \mathcal{M}^{I \to N,\overline{R}}_{\mathrm{virt}} \prod^r_{j=1} \mathcal{M}^{j \to R_j}_{\mathrm{LO}} \nonumber
        \end{aligned} \\
        &\quad{} +
        \begin{aligned}[t]
                &  \mathcal{M}^{I \to N,\overline{R}}_{\mathrm{LO}} \sum^r_{k=1} \mathcal{M}^{k \to R_k}_{\mathrm{virt}}
                           \prod^r_{j \neq k} \mathcal{M}^{j \to R_j}_{\mathrm{LO}} \Bigg]_{ \left\{ \overline{k}^2_l \to {\widehat{\overline{k}}}^2_l = M^2_l \right\}_{l \in \overline{R}} } . \\
        \end{aligned}
\end{align}
The propagator of the resonant particle $i$ is $K_i = \overline{k}^2_i
- \overline{M}^2_i$, where $\overline{M}^2_i = M^2_i - \text{i} M_i
\Gamma_i$ is its complex mass squared.  The on-shell projection is
applied everywhere in the matrix element but in the resonant
propagators $K_i$ and is indicated by $\left\{ \overline{k}^2_l \to
  {\widehat{\overline{k}}}^2_l = M^2_l \right\}$.  The indices $I$,
$\overline{R}$, $R_i$ and $N$ denote the ensembles of initial
particles, resonant particles, decay products of the resonant particle
$i$, and the final-state particles not resulting from the decay of a
resonant particle.  Finally $\lambda_i$ represents the polarisations
of the resonance $i$.  Note that these virtual factorisable corrections
have been obtained via the computer code \recola, which allows to
select resonant contributions at both LO and NLO.

When taking the on-shell limit of the momenta of the resonant
particles, artificial IR singularities related to charged resonances are
introduced.  These artificial IR divergences are cancelled when
including non-factorisable corrections.  The non-factorisable
corrections result from diagrams that do not split into production and
decay parts by cutting only the resonant lines but also include
contributions from factorisable diagrams \cite{Denner:2000bj}.  The
latter are obtained by taking the factorisable diagrams, where the IR
singularities related to the resonant particles are regularised by the
finite decay widths, and subtracting these contributions for zero
decay width, which contains the artificial IR-divergent piece
mentioned previously.  The non-factorisable corrections can be written
in the form
\eq{2 \mathrm{Re} \left\{ \mathcal{M}^*_{\mathrm{LO,PA}} \mathcal{M}_{\mathrm{virt,nfact,PA}} \right\} = |\mathcal{M}_{\mathrm{LO,PA}}|^2 \delta_{\mathrm{nfact}} ,}
where $\delta_{\mathrm{nfact}}$ is a function of scalar integrals (computed using the \collier library) whose expression can be found in \citeres{Dittmaier:2015bfe,Denner:2016jyo}.

As explained previously, the ${ q \bar{q}}$ channels have two kinds of
virtual NLO contributions connected by IR divergences: one-loop
amplitudes of order $\order{\alphas\alpha^{7/2}}$ interfered with tree
amplitudes of order $\order{\alphas\alpha^{5/2}}$ and the interference
of one-loop amplitudes of order $\order{\alphas^2\alpha^{5/2}}$ with
tree amplitudes of order $\order{\alpha^{7/2}}$.
Thus, the DPA must be applied to both types of virtual corrections.

\subsection{Validation}
\label{sec:Validations}

The hadronic cross section for the full LO process as well as the
photon-induced channels for dynamical scale have been checked against
the computer code \madgraph~\cite{Alwall:2014hca}.  Agreement within
the statistical error has been obtained.  At the squared
matrix-element level, we have checked all the tree-level contributions
against \madgraph.  For Born, computing 10000 random phase-space
points, we have found an agreement of at least 11 digits for $99\%$ of
the points, whereas for the real contribution, the two codes agree for
$99\%$ of the points within at least 8 digits.

To verify the IR and ultra-violet (UV) finiteness, the cross section
has been computed for different IR and UV regulators, respectively.
In order to check the proper subtraction of IR divergences the
$\alpha$ parameter\footnote{The results presented in this article are obtained for $\alpha= 10^{-2}$.} has been varied from $10^{-2}$ to
1.  The parameter $\alpha$ can restrict the phase space for the dipole
subtraction terms to the vicinity of the singular regions
\cite{Nagy:1998bb} and thus improves numerical stability.  We have
also checked a Ward identity for the gluon-induced virtual amplitude, substituting in the one-loop amplitude the polarisation
vector of one of the initial-state gluons by its momentum normalised
to its energy, as $\epsilon^\mu_\Pg \to p^\mu_\Pg/p^0_\Pg$.  Looking
at the cumulative fraction of events with
$\Re \left[\mathcal{M}^*_0(\epsilon_\Pg)\mathcal{M}_1(\epsilon_\Pg\to
p_\Pg/p^0_\Pg)\right]/\Re\left[\mathcal{M}^*_0(\epsilon_\Pg)\mathcal{M}_1(\epsilon_\Pg)\right]$
larger than given values, we have found comparably good results to
those of \citeres{Denner:2015yca,Denner:2016jyo}.  We remark that
the Ward identity is completely spoiled if we omit the contributions of 9-point functions.
For the gluon-induced channel, the full virtual contribution to the integrated cross section is $0.0555(6)$.
Without 9-point functions, it is $0.381(4)$, while the contribution of 9-point functions alone is $-0.322(5)$.
This indicates that the 9-point functions yield a sizeable contributions to the cross section.
Finally, to check the virtual corrections, two DPAs
have been used and agree very well with the full computation (see
below).

Concerning the QCD corrections, we have simply reproduced
representative contributions of \citere{Denner:2015yca} as this
computation has undergone already numerous checks.
Since we use the same Monte Carlo program and have access to the splitting of every single contribution, we have been able to check each of them in detail.

\section{Numerical Results}
\label{sec:results}

\subsection{Input parameters and event selection}
\label{ssec:InputParameters}

In this section, we present predictions for the LHC operating at a centre-of-mass energy $\sqrt{s}=13\TeV$.
In particular, the integrated cross section and differential distributions including NLO EW corrections are reported.
The PDFs have been incorporated through LHAPDF
6.1.5~\cite{Andersen:2014efa,Buckley:2014ana}.  Specifically, the
$\mathrm{LUXqed}\_\mathrm{plus}\_\mathrm{PDF4LHC15}\_\mathrm{nnlo}\_100$
set~\cite{Manohar:2016nzj} has been used for all LO and NLO
results.  It is next-to-next-to-leading-order (NNLO) accurate in QCD
and includes all terms of order $\alpha L \left(\alpha_s L\right)^n$,
$\alpha \left(\alpha_s L\right)^n$, and $\alpha^2 L^2 \left(\alpha_s
  L\right)^n$, where $L = \ln(\mu^2 / m^2_{\rm p}) $, $\mu$ is the
renormalisation scale, and $m_{\rm p}$ is the proton mass.  For the
QCD partons, this PDF set is based on
\citeres{Butterworth:2015oua,Dulat:2015mca,Harland-Lang:2014zoa,Ball:2014uwa,Gao:2013bia,Carrazza:2015aoa}.
Moreover, it features the inclusion of an $\overline{\text{MS}}$ photon
PDF needed for the photon-initiated contributions.  For the
renormalisation and factorisation scale, the following dynamical scale
has been chosen \cite{Denner:2015yca}:
\begin{equation}\label{eqn:dynScale}
  \mu_\text{dyn} = \left(m_\text{T,t} m_{\text{T},\overline{\text{t}}} m_\text{T,H} \right)^{\frac13} \quad \text{with} \quad m_\text{T} = \sqrt{m^2 + p_{\text{T}}^2} .
  \end{equation}
Note that contributions for bottom-quark PDFs have been neglected.

The $G_\mu$ scheme \cite{Denner:2000bj} has been used where $\alpha$ is obtained from the Fermi constant,
\begin{equation}\label{eqn:FermiConstant}
  \alpha = \frac{\sqrt{2}}{\pi} G_\mu \MW^2 \left( 1 - \frac{\MW^2}{\MZ^2} \right)
  \qquad \text{with}  \qquad   \GF    = 1.16637\times 10^{-5}
\change{\GeV^{-2}},
\end{equation}
leading to $\alpha = 0.0075553105 \ldots$ .

The numerical values for the masses and widths used in this computation read \cite{Beringer:1900zz}:
\begin{alignat}{2} \label{eqn:ParticleMassesAndWidths}
                 \Mt   &=  173.34\GeV,       & \quad \quad \quad  \Gt^\text{NLO} &= 1.36918\ldots\GeV,  \nonumber \\
                \MZOS &=  91.1876\GeV,      & \GZOS &= 2.4952\GeV, \nonumber \\
                \MWOS &=  80.385\GeV,       & \GWOS &= 2.085\GeV,  \nonumber \\
                M_{\rm H} &=  125.0\GeV.       & 
\end{alignat}
The masses and widths of all other quarks and leptons have been
neglected.  The effect of a finite bottom-quark mass on the cross
section has been found to be below the per-cent level in our set-up
\cite{Denner:2015yca}.  The values of the top-quark widths (LO and
NLO) have been taken from \citere{Basso:2015gca}, where both EW and
QCD NLO corrections for massive bottom quarks have been calculated.
The effect of the bottom-quark mass on the top-quark width has been
found to be negligible compared to the integration errors on the cross
section \cite{Denner:2016jyo}.  Note that the Higgs mass stated above
is not the one used in \citere{Basso:2015gca} but the one
recommended by the Higgs Cross Section Working
Group~\cite{deFlorian:2016spz} (such a change has a negligible impact
on the NLO EW corrections).  Finally, in this article we have used two
different top-quark widths for the LO contributions.  When presenting
results for the NLO EW corrections (and their comparison to the two
DPAs) the width $\Gt^\text{NLO,QCD} = 1.35029\GeV$ which features NLO QCD
corrections is used at LO.
This yields relative EW correction that can be directly compared to EW corrections in computations with on-shell top quarks.
On the other hand, when presenting
the combined NLO EW and QCD results, the width used for the LO
predictions, $\Gt^\text{LO} = 1.449582\GeV$, does not incorporate any NLO
corrections.

The pole values for the gauge-boson ($V=\PW,\PZ$) parameters are
obtained from the measured on-shell (OS) values of the masses and
widths, according to \citere{Bardin:1988xt},
\newcommand{\MVOS}{\ensuremath{M_V^\text{OS}}\xspace}%
\newcommand{\GVOS}{\ensuremath{\Gamma_V^\text{OS}}\xspace}%
\begin{equation}
        M_V = \MVOS/\sqrt{1+(\GVOS/\MVOS)^2}\,,\qquad  \Gamma_V = \GVOS/\sqrt{1+(\GVOS/\MVOS)^2}.
\end{equation}

The anti-$k_\text{T}$ algorithm \cite{Cacciari:2008gp}, is used to
cluster QCD partons and photons into jets as well as photons with
light charged particles, using a jet-resolution parameter $R=0.4$.  In
the rapidity--azimuthal-angle plane, the distance between two
particles $i$ and $j$ is defined as
\begin{equation}\label{eqn:DeltaR}
        R_{ij} = \sqrt{(\Delta \phi_{ij})^2+(y_i-y_j)^2},
\end{equation}
where $\Delta \phi_{ij}$ is the azimuthal-angle difference.  The
rapidity of jet $i$ reads $y_i=\frac{1}{2}\ln \frac{E+p_z}{E-p_z}$, 
where $E$ is the energy of the jet and $p_z$ the component of its
momentum along the beam axis.  Only final-state quarks, gluons, and
charged fermions with rapidity $|y|<5$ are clustered into IR-safe
objects.

After recombination, standard event selections are applied on the
transverse momenta and rapidities of charged leptons and b~jets,
missing transverse momentum and rapidity--azimuthal-angle distance
between bottom jets.  The Higgs boson is not included in the event
selection.  In the final state, two bottom jets and two charged
leptons are required, and the following event selection is applied:
\begin{alignat}{5} \label{eqn:cuts}
                 \text{bottom jets:}                     && \qquad \ptsub{\Pb}         &>  25\GeV,  & \qquad |y_\Pb|   &< 2.5, & \nonumber \\
                \text{charged lepton:}              && \ptsub{\Pl}         &>  20\GeV,  & |y_{\Pl}| &< 2.5, &\nonumber \\
                \text{missing transverse momentum:} && \ptsub{\text{miss}} &>  20\GeV,                      &\nonumber \\
                \text{bottom-jet--bottom-jet distance:}       && \Delta R_{\Pb\Pb}   &> 0.4.                          &
\end{alignat}

\subsection{EW corrections to integrated cross section}
\label{ssec:IntegratedCrossSection}

In this section the results for the integrated cross section for the
LHC at a centre-of-mass energy of $\sqrt{s}=13\TeV$ are presented.
The different contributions are summarised in
\refta{table:results_summary}.  These predictions are made for the
input parameters given in
Eqs.~\eqref{eqn:dynScale}--\eqref{eqn:ParticleMassesAndWidths} and the
event selections defined in Eq.~\eqref{eqn:cuts}.  We consider LO
contributions of the order $\order{\alphas^2\alpha^{5}}$, while the EW
NLO corrections arise at order $\order{\alphas^2\alpha^{6}}$.  The
${\rm \gamma g}$ contributions being of the order $\order{\alphas
  \alpha^{5}}$, have not been included in the total cross section and are shown only for reference.

\begin{table}
\begin{center} 
\begin{tabular}{ c  c  c   c }
 Ch. & $\sigma_{\rm LO}$ [fb] & $\sigma_{\rm NLO \; EW}$ [fb] & $\delta$ [$ \% $]\\
  \hline\hline
$\Pg \Pg$        &  $2.0116(1)$ & $2.020(1) $ & $+0.42$ \\
${ q \bar{q}}$      &  $0.84860(5)$ & $0.8454(6) $ & $-0.38$ \\
${\rm g} q(/\bar{q})$    &                                   & $0.00007(2)  $ &          \\
  \hline
${\rm \gamma g}$  &  \multicolumn{2}{c}{$0.02178(1)$}   &      \\
\hline\hline
$\Pp\Pp$         &  $ 2.8602(1) $ & $2.866(1) $ & $+0.20$ \\
  \hline
\end{tabular}
\end{center}
        \caption[Composition of the integrated cross section]{\label{table:results_summary}
                Contributions to the integrated cross section for $\Pp\Pp \to \Pe^+\nu_\Pe \mu^- \bar{\nu}_\mu \Pb \bar{\Pb} \PH (\gamma / \Pj)$ at the LHC at a centre-of-mass energy of $\sqrt{s}=13\TeV$.
                The quark--antiquark contributions include $ q=\Pu,\Pd,\Pc,\Ps$.
                The contribution originating from the real radiation of a quark or an antiquark is denoted by ${\rm g} q(/\bar{q})$.
                The total cross section (denoted by $\Pp\Pp$) does not include the photon-induced channel (denoted by ${\rm \gamma g}$).
                The definition of the relative corrections reads $\delta = \sigma_{\rm NLO \; EW}/\sigma_{\rm LO}$.
                The integration errors of the last digits are given in parentheses.}
\end{table}
Due to the enhanced gluon PDF, the gluon--gluon-initiated channel is
dominating.  The ${ q \bar{q}}$ channels including ${
  q}=\Pu,\Pd,\Pc,\Ps$ make up $30\%$ of the total cross section.  This
is in contrast to the production of a pair of off-shell top quarks
\cite{Denner:2016jyo} where the ${ q \bar{q}}$ channels account for
only $12\%$ of the total cross section.  While the corrections to the
gluon--gluon-initiated channel are positive, those to the ${ q
  \bar{q}}$ channels are negative amounting $+0.42\%$ and $-0.38\%$,
respectively.  Moreover, the ${\rm g} q/\bar{q}$ channels give a
negligible contribution to the total cross section.  In the end, for
the full hadronic process, the EW corrections contribute $0.20 \%$.
Note that the photon contributions account for $0.76\%$ of the NLO
cross section.

In \citere{Frixione:2015zaa}, EW corrections for the production of
a Higgs boson in association with on-shell top quarks have been
reported.  In the $G_{\mu}$ scheme, they amount to $+1.8\%$ and
comprise photon contributions of $0.2\%$.  In our calculation the
photon-induced contributions have not been included in the definition
of the NLO EW corrections.  Accounting for this effect, the two
computations agree within $1.5\%$.  Note that cuts on the final state
are applied in our computation which also includes off-shell and
non-resonant contributions.  By definition, these are not taken into
account in on-shell computations.

Following \citere{Denner:2015yca}, the bottom quarks have been considered massless and the bottom-quark PDF has been neglected.
This is justified by the fact that they contribute at the per-mille level.

To conclude, the EW corrections are below the per-cent level for the
integrated cross section.  For differential distributions, on the
other hand, the EW corrections have a larger impact (see below).

\subsection{EW corrections to differential distributions}
\label{ssec:DifferentialDistributions}

Turning to differential distributions, two plots are shown for each
observable.  The upper panels display the LO and NLO EW predictions,
while in the lower panels the relative corrections $\delta =
\sigma_\text{NLO EW} / \sigma_\text{LO} - 1$ are shown in per cent.
For reference, the ${\rm \gamma g}$ contribution is also displayed as
$\delta_{\rm \gamma g} = \sigma_{\rm \gamma g} / \sigma_\text{LO}$ and
labelled {\em{photon}}.  As opposed to the distributions shown in
\citere{Denner:2016jyo}, we have restricted the range of the
transverse-momentum distributions to $400\GeV$.  As the process is
just about to be measured, it is unlikely that the experimental
collaborations will be able to probe the very tail of the
distributions in a near future.
\begin{figure}
        \setlength{\parskip}{-10pt}
        \begin{subfigure}{0.49\textwidth}
                \subcaption{}
                \includegraphics[width=\textwidth]{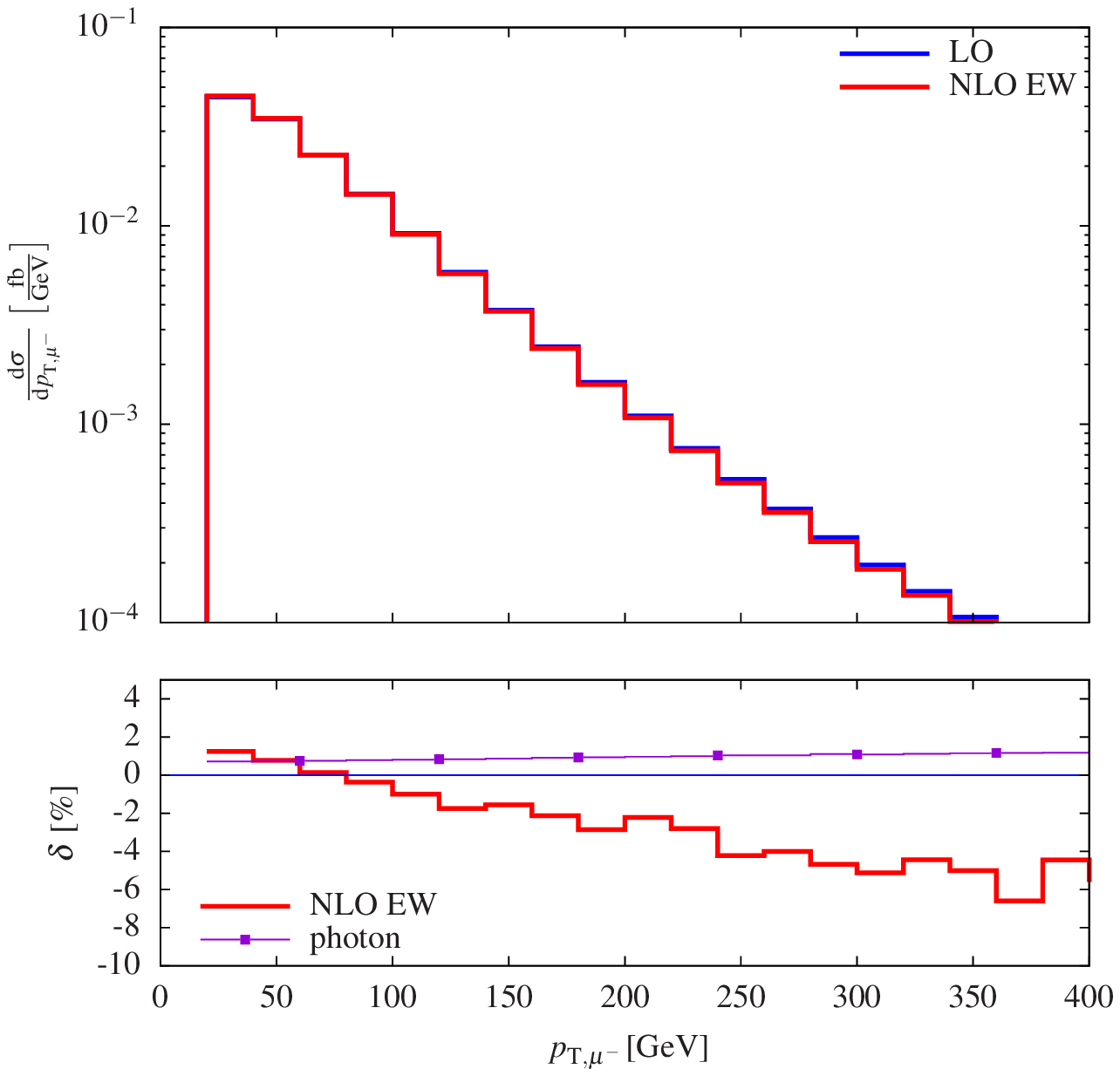}
                \label{plot:transverse_momentum_muon}
        \end{subfigure}
        \hfill
        \begin{subfigure}{0.49\textwidth}
                \subcaption{}
                \includegraphics[width=\textwidth]{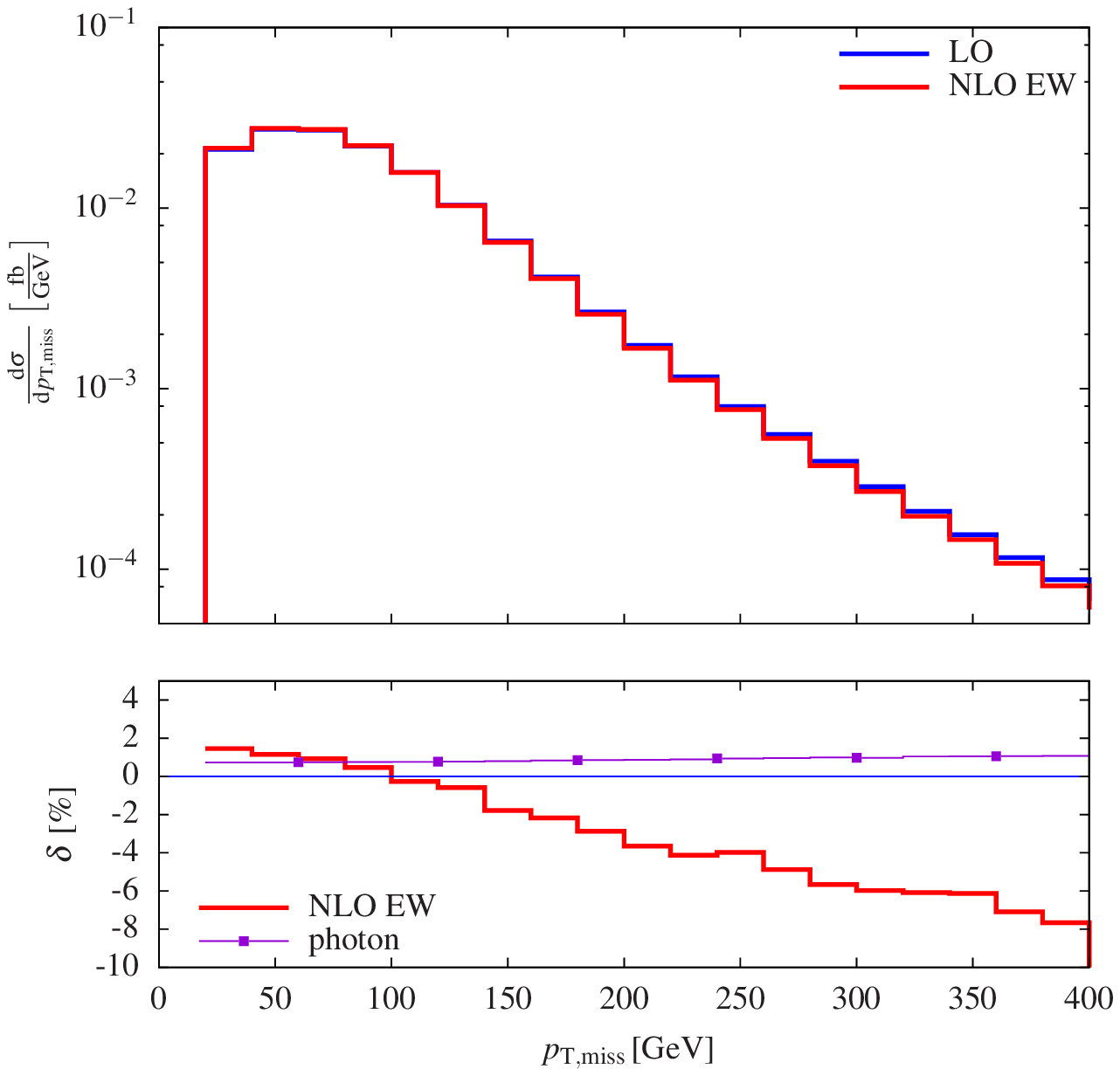}
                \label{plot:transverse_momentum_truth_missing} 
        \end{subfigure}
        
        \begin{subfigure}{0.49\textwidth}
                \subcaption{}
                \includegraphics[width=\textwidth]{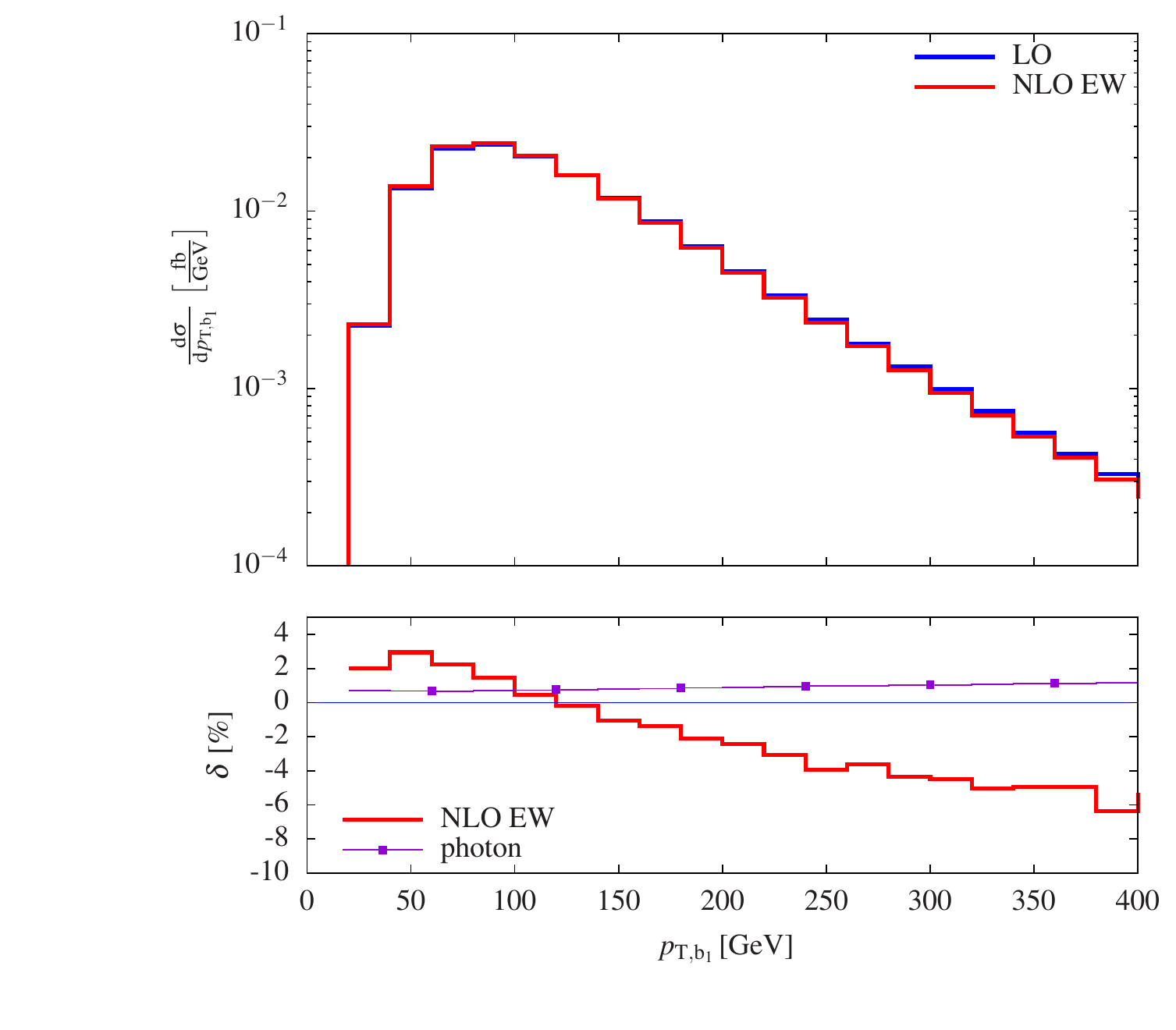}
                \label{plot:transverse_momentum_b1}
        \end{subfigure}
        \hfill
        \begin{subfigure}{0.49\textwidth}
                \subcaption{}
                \includegraphics[width=\textwidth]{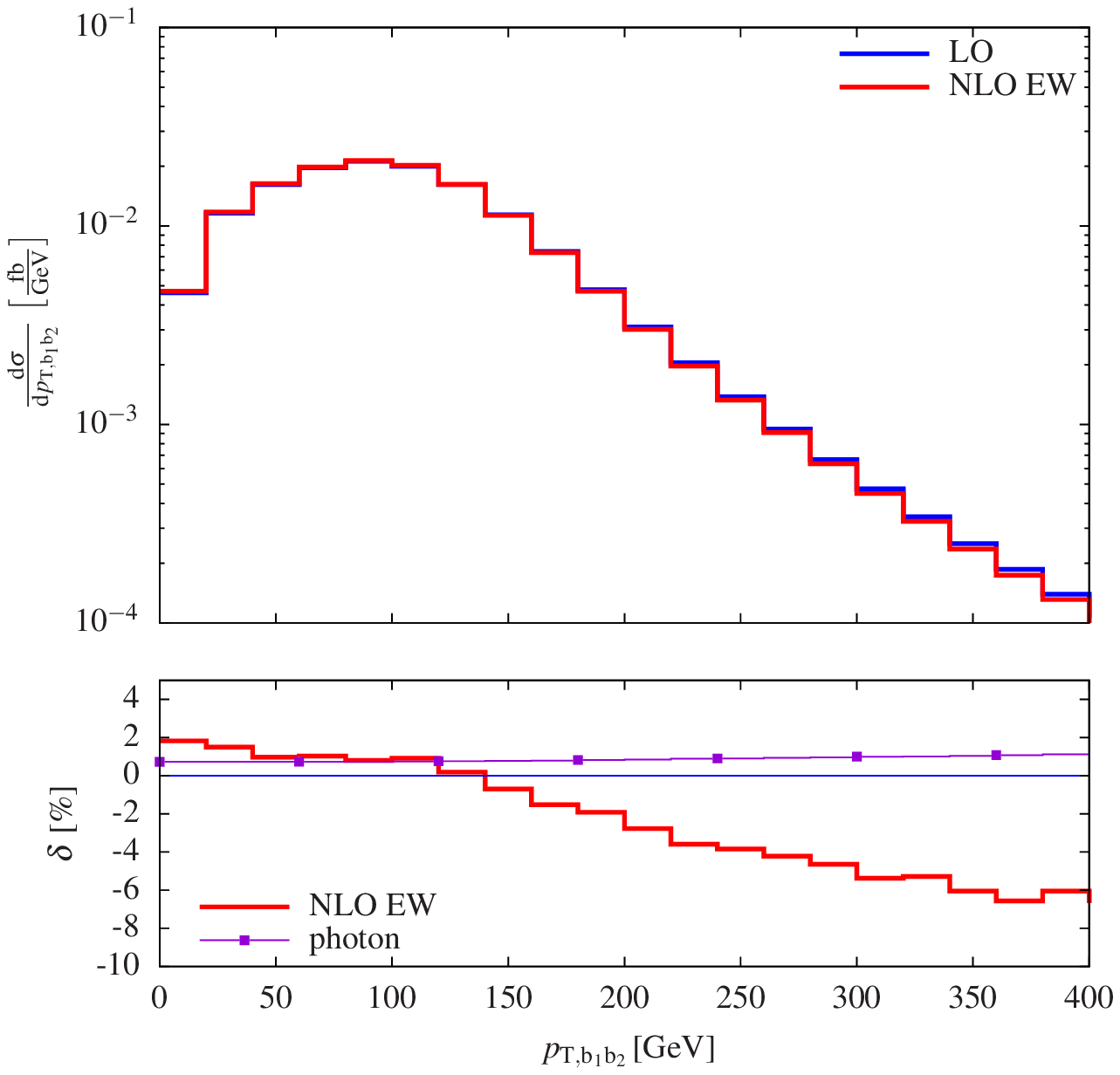}
                \label{plot:transverse_momentum_bb12}
        \end{subfigure}

        \begin{subfigure}{0.49\textwidth}
                \subcaption{}
                \includegraphics[width=\textwidth]{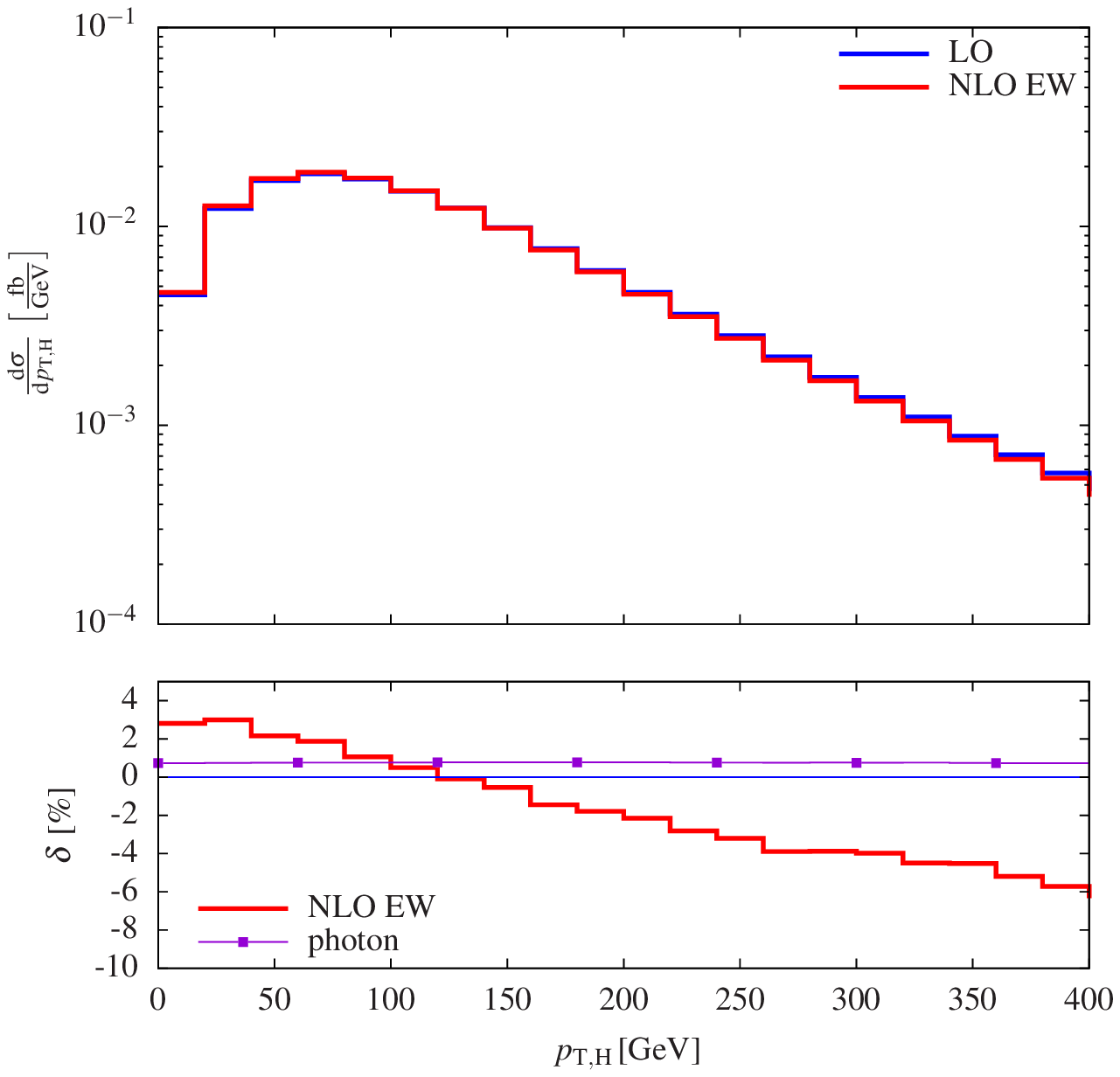}
                \label{plot:transverse_momentum_higgs}
        \end{subfigure}
        \hfill
        \begin{subfigure}{0.49\textwidth}
                \subcaption{}
                \includegraphics[width=\textwidth]{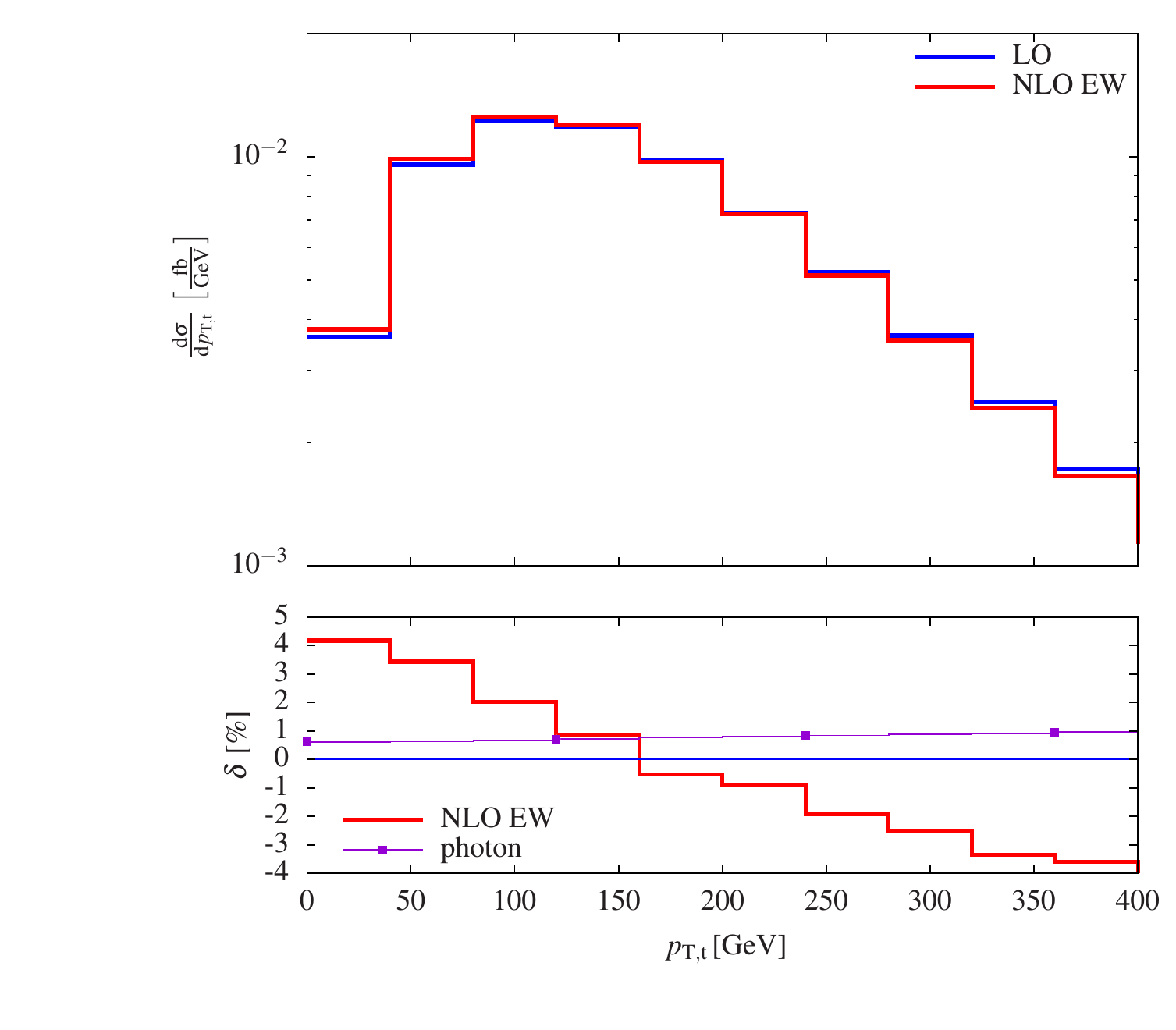}
                \label{plot:transverse_momentum_top}
        \end{subfigure}
        
        \vspace*{-3ex}
        \caption{\label{fig:transverse_momentum_distributions}%
                Transverse-momentum distributions at the LHC running at a centre-of-mass energy $\sqrt{s}=13\TeV$: 
                \subref{plot:transverse_momentum_muon} for the muon~(upper left), %
                \subref{plot:transverse_momentum_truth_missing} for the missing momentum~(upper right), %
                \subref{plot:transverse_momentum_b1} for the harder bottom jet~(middle left), %
                \subref{plot:transverse_momentum_bb12} for the bottom-jet pair~(middle right), %
                \subref{plot:transverse_momentum_higgs} for the Higgs boson~(lower left), and %
                \subref{plot:transverse_momentum_top} for the reconstructed top quark~(lower right).
                In the lower panels, the relative NLO EW corrections $\delta = \sigma_{\text{NLO EW}} / \sigma_\text{LO} - 1$ and the relative photon-induced
                contributions $\delta = \sigma_{ \gamma g } /
                \sigma_\text{LO} $ in per cent are shown.  }
\end{figure}%

In Figure~\ref{fig:transverse_momentum_distributions}, a selection of
transverse-momentum distributions is shown.  In all distributions, the
effects of the Sudakov logarithms at high transverse momenta are
clearly visible.  In general, the corrections vary between $1 \%$ and
$4\%$ for transverse momenta below 50\GeV and grow negative towards
high transverse momenta.  At the end of the range shown ($400\GeV$),
the EW corrections reach up to $-8\%$.

This is exemplified in Figure~\ref{plot:transverse_momentum_muon}, where the distribution of the muon transverse momentum is presented.
We have checked that the negative corrections for large transverse
momenta are solely driven by the virtual contributions containing the
Sudakov logarithms (not shown separately in the plots).  The
distribution in the missing transverse momentum in
\reffi{plot:transverse_momentum_truth_missing} features the largest EW
corrections which amount to $-8\%$ at $400\GeV$.  The missing momentum
is defined as the sum of the transverse momenta of the two neutrinos,
\emph{i.e.}~$\ptsub{\text{miss}}=\abs{\vec{p}_{\text{T},\nu_{\Pe}}+\vec{p}_{\text{T},\bar{\nu}_{\mu}}}$.
In \reffis{plot:transverse_momentum_b1} and
\ref{plot:transverse_momentum_bb12} the transverse momentum of the
harder bottom quark (according to $\pt$ ordering) and the transverse
momentum of the bottom-quark pair are displayed, respectively.  Both
observables receive corrections reaching $-6\%$ at $400 \GeV$ and show
similar behaviours.  \reffi{plot:transverse_momentum_higgs} shows the distribution in the
transverse momentum of the produced Higgs boson.  At zero transverse
momentum, the EW corrections amount to about $3\%$ to reach $-6\%$ for
a Higgs boson with a transverse momentum of $400\GeV$.  Finally, the
transverse momentum of the reconstructed top quark is displayed in
\reffi{plot:transverse_momentum_top}.  There, the EW corrections start
at the level of $+4\%$ below $50 \GeV$ to reach $-4\%$ at $400 \GeV$.

In contrast to previous computations for top-quark-pair production
\cite{Pagani:2016caq,Denner:2016jyo}, the photon-induced contributions
are stable over the whole range.  The reason is that in the previous
computations, the
$\mathrm{NNPDF23}\_\mathrm{nlo}\_\mathrm{as}\_0119\_\mathrm{qed}$
set~\cite{Ball:2013hta,Carrazza:2013bra,Carrazza:2013wua} has been
used, while for the present computation the
$\mathrm{LUXqed}\_\mathrm{plus}\_\mathrm{PDF4LHC15}\_\mathrm{nnlo}\_100$
set~\cite{Manohar:2016nzj} is employed.  The latter features small photon
PDF contributions and smaller errors.  In particular, in the
high-energy tail of distributions, the photon-induced contributions
stay below one per cent while they were reaching $10\%$ with previous
PDF sets.
 
\begin{figure}
        \setlength{\parskip}{-10pt}
        \begin{subfigure}{0.49\textwidth}
                \subcaption{}
                \includegraphics[width=\textwidth]{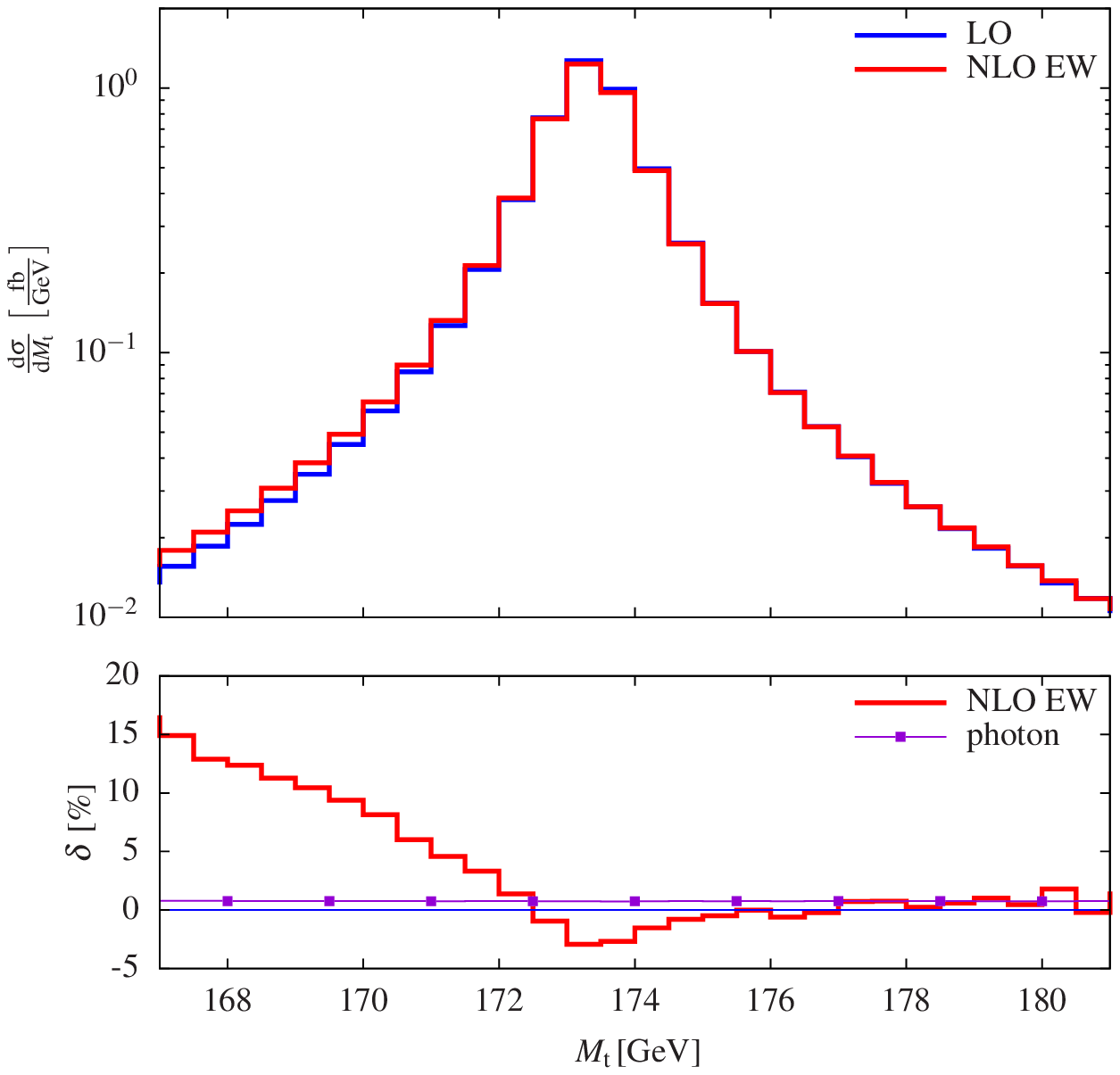}
                \label{plot:invariant_mass_truth_top}
        \end{subfigure}
        \hfill
        \begin{subfigure}{0.49\textwidth}
                \subcaption{}
                \includegraphics[width=\textwidth]{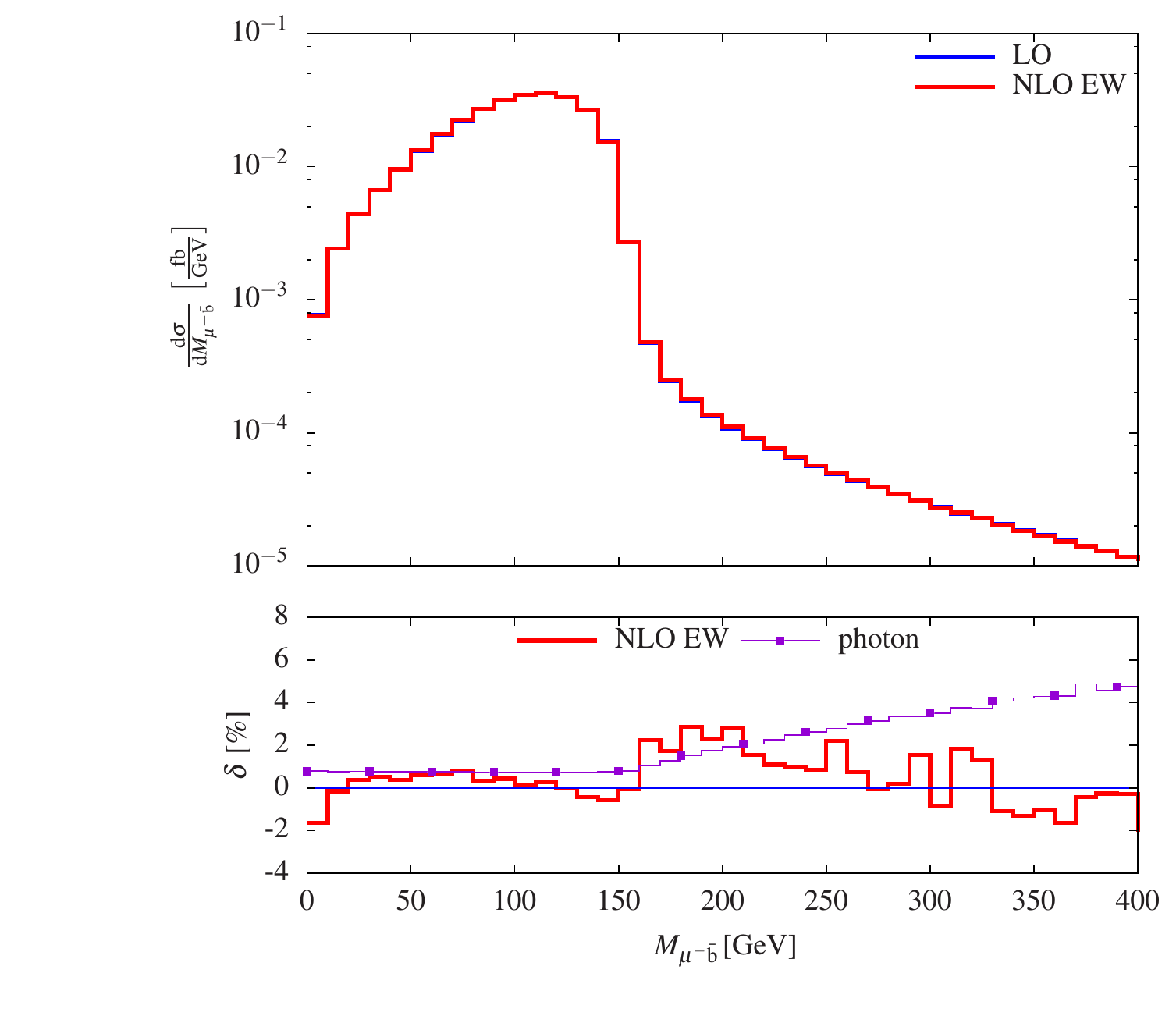}
                \label{plot:invariant_mass_truth_mubx}
        \end{subfigure}
        \begin{subfigure}{0.49\textwidth}
                \subcaption{}
                \includegraphics[width=\textwidth]{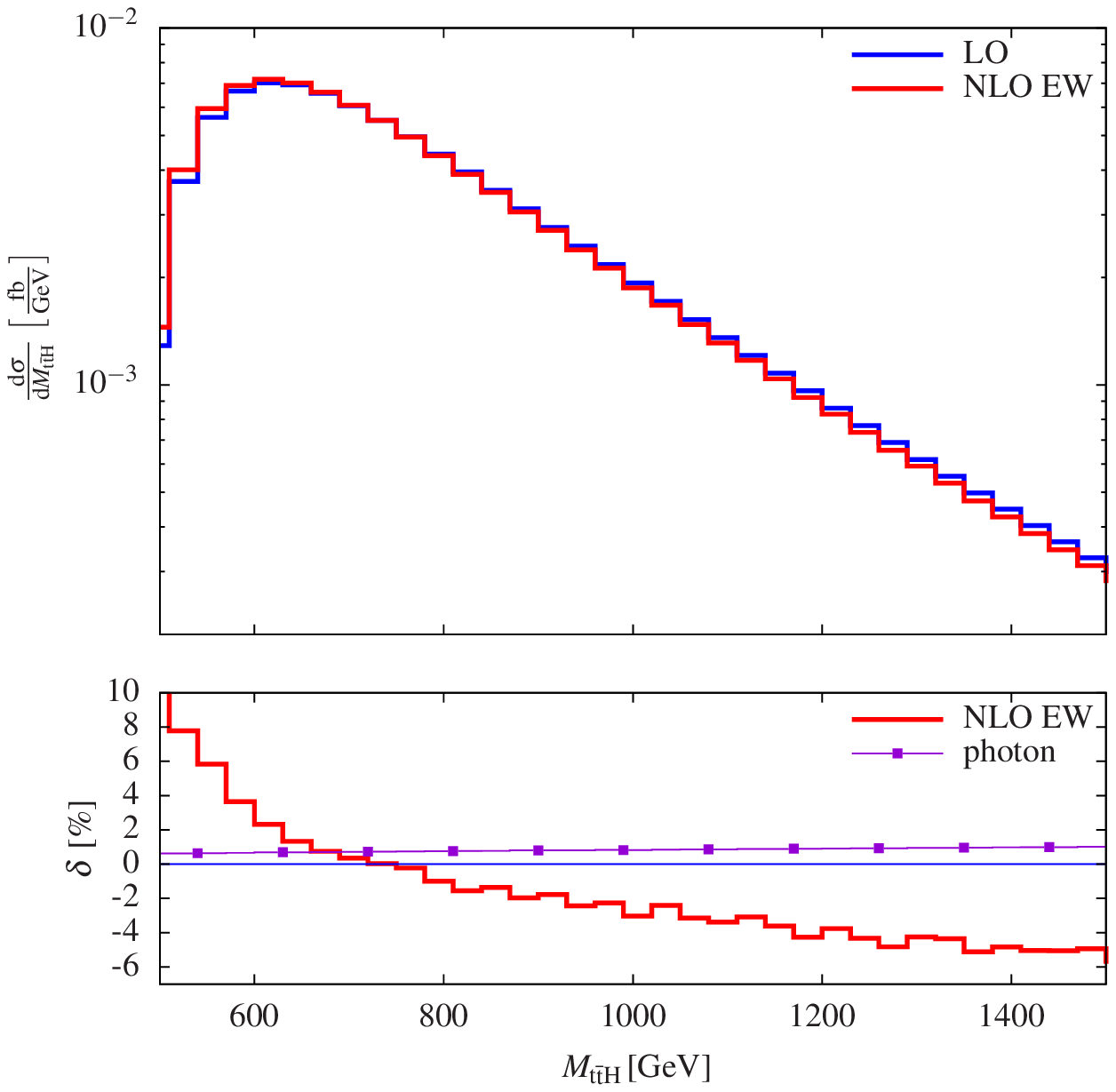}
                \label{plot:invariant_mass_truth_ttxh}
        \end{subfigure}
        \hfill
        \begin{subfigure}{0.49\textwidth}
                \subcaption{}
                \includegraphics[width=\textwidth]{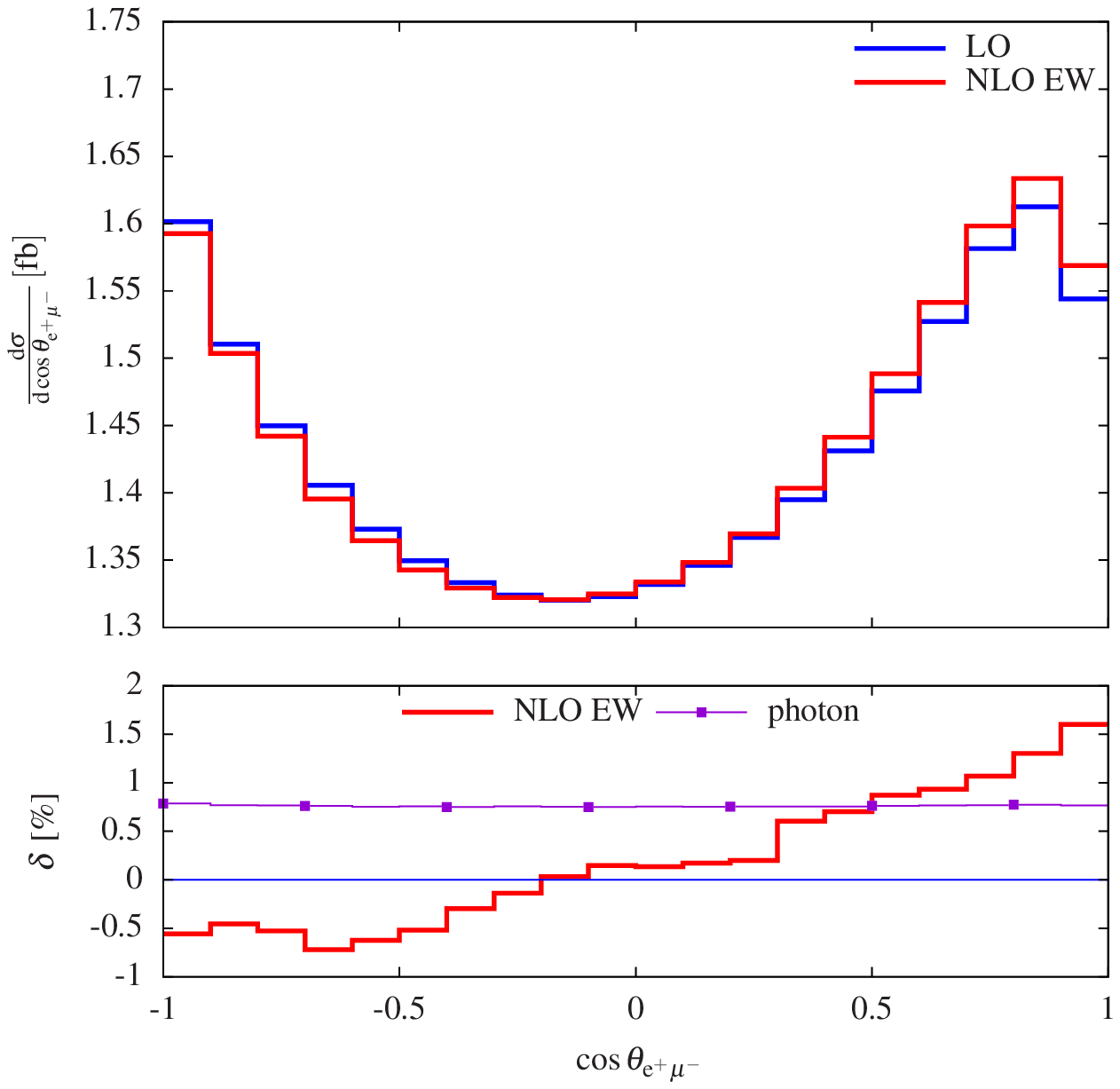}
                \label{plot:cosine_angle_separation_epmu} 
        \end{subfigure}

                \begin{subfigure}{0.49\textwidth}
                \subcaption{}
                \includegraphics[width=\textwidth]{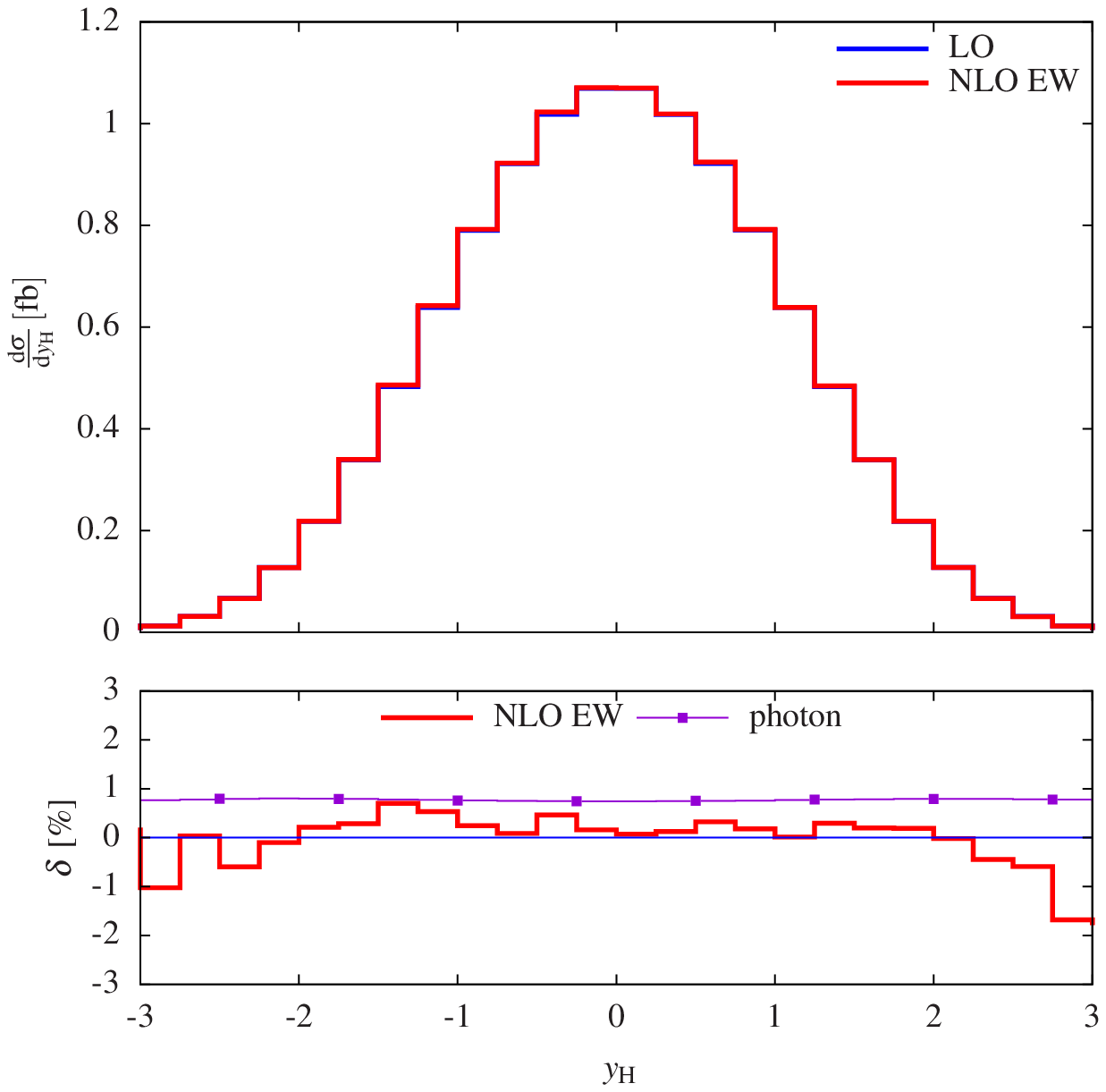}
                \label{plot:rapidity_higgs}
        \end{subfigure}
        \hfill
        \begin{subfigure}{0.49\textwidth}
                \subcaption{}
                \includegraphics[width=\textwidth]{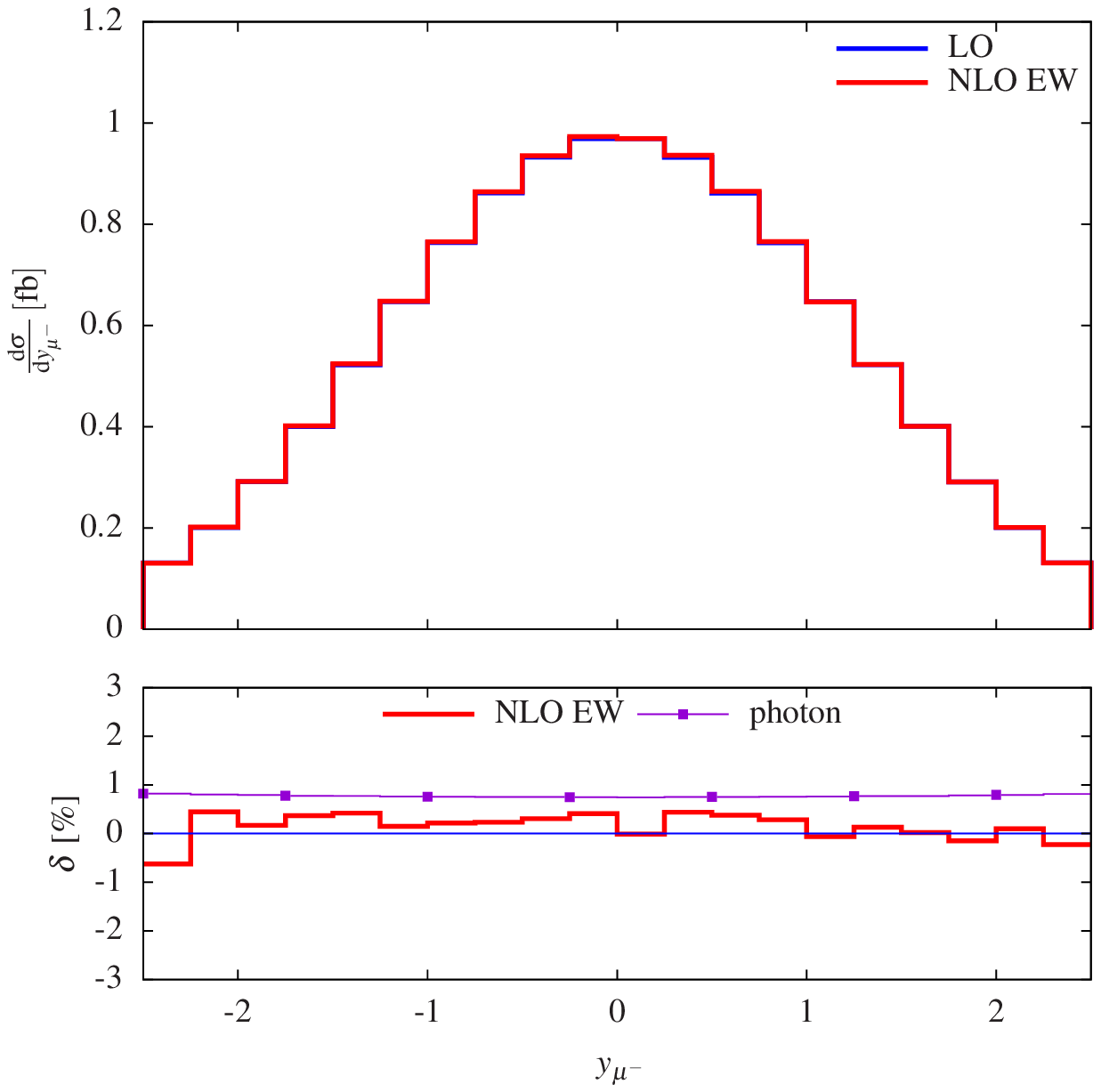}
                \label{plot:rapidity_muon} 
        \end{subfigure}
        
        \vspace*{-3ex}
        \caption{\label{fig:various_differential_distributions}%
                Distributions at the LHC running at a centre-of-mass energy $\sqrt{s}=13\TeV$: 
                \subref{plot:invariant_mass_truth_top}~invariant mass of the reconstructed top quark~(upper left),
                \subref{plot:invariant_mass_truth_mubx}~invariant mass of the $\mu^- \bar{\Pb}$ system~(upper right),
                \subref{plot:invariant_mass_truth_ttxh}~invariant mass of the reconstructed $\Pt\bar{\Pt}\PH$ system~(middle left), 
                \subref{plot:cosine_angle_separation_epmu}~cosine of the angle between the positron and the muon~(middle right),
                \subref{plot:rapidity_higgs}~rapidity of the Higgs boson~(lower left), 
                and \subref{plot:rapidity_muon}~rapidity of the muon~(lower right).
                In the lower panels, the relative NLO EW corrections $\delta = \sigma_{\text{NLO EW}} / \sigma_\text{LO} - 1$ and the relative photon-induced
                contributions $\delta = \sigma_{ \gamma g } /
                \sigma_\text{LO} $ in per cent are shown. }
\end{figure}%

We turn to invariant-mass and angular distributions.  As it is well
known \cite{Denner:2012yc,Denner:2015yca,Denner:2016jyo}, the
corrections to the reconstructed invariant top mass in
\reffi{plot:invariant_mass_truth_top} display a radiative tail below
the top-mass threshold.  This is due to photons which have not been
reconstructed with final-state particles.  At $167\GeV$, the
corrections reach $+15\%$ while at $181\GeV$ they are practically
zero.  At the nominal top mass, the EW corrections amount to $-2\%$.

The distribution in the invariant mass of the muon--antibottom system
displayed in \reffi{plot:invariant_mass_truth_mubx} is particularly
interesting.
The threshold $ M^2_{\rm t} - M^2_{\rm W} \simeq (154\GeV)^2$ marks
the transition between the on-shell and off-shell production of the
top-antitop quark pair
\cite{Denner:2012yc,Denner:2015yca,Denner:2016jyo}.  Below and above
this point, the EW corrections vary between $-1\%$ and $+1\%$ while
they increase to about $+2\%$ at threshold.  One also notices that the
photon-induced contributions are constant below threshold at the level
of $1\%$ but start to increase above threshold to reach $+5\%$ at
$400\GeV$.

The distribution in Figure~\ref{plot:invariant_mass_truth_ttxh} displays the invariant mass of the reconstructed t$\bar{\rm t}$H system.
It includes all the particles produced and thus shows the EW correction as a function of the partonic centre-of-mass energy.
At $500\GeV$, \emph{i.e.}~just above the threshold $2 \Mt + M_{\rm H} \simeq 472 \GeV $, the corrections are at the level of $10\%$ and go down to $-5\%$ at $1500\GeV$.

The distribution in the cosine of the angle between the positron and the muon is displayed in Figure~\ref{plot:cosine_angle_separation_epmu}.
As for QCD corrections (with a larger amplitude), the relative EW ones are increasing between $\cos \theta_{e^+ \mu^-} = -1$ and $\cos \theta_{e^+ \mu^-} = 1$ from $-0.5\%$ to $+1.5\%$.

Finally, we show the rapidity distribution of both the Higgs boson and
the muon in Figures~\ref{plot:rapidity_higgs} and
\ref{plot:rapidity_muon}.  The other rapidity
distributions also do not display any particular EW corrections, and no
variation is observed over the whole rapidity range.

For the observables involving the reconstructed top quarks and/or the
final-state Higgs boson, qualitatively we have found similar results
than the ones presented in \citere{Frixione:2015zaa}.  As for the
total cross section, a quantitative comparison is not possible due to
the event selection applied in our off-shell calculation.

\subsection{Comparison to the double-pole approximations}
\label{sec:ComparisonDPA}

In this section a study of two different DPAs for the off-shell
production of top-quark pairs in association with a Higgs boson is
presented.  In particular, by comparing them to the full calculation
at the level of integrated cross section and differential
distributions, we can infer the quality of these two approximations.
The first approximation requires two resonant top quarks, while for
the second two resonant W bosons are demanded.

\subsubsection*{Integrated cross section}

To start, we report results at LO for the integrated cross section for
both the $\Pg \Pg$ and ${ q \bar{q}}$ channels in
\refta{table:LO_DPA_results_summary}.  While the WW DPA agrees with
the full LO result within one per cent, the tt DPA only agrees at the
level of $1.5$--$2\%$.  The WW DPA approximates the full calculation
better as it comprises most diagrams of the full process.  In
particular, it contains all the doubly and singly top-quark resonant
contributions as well as some non-resonant contributions while by
definition the tt DPA only contains the doubly-resonant top-quark
contributions.  Nonetheless, the agreement matches the order of
magnitude $\Gamma/M$ expected for a DPA.  Since the ${\rm \gamma g}$
channels contribute below the per-cent level and the associated
QCD corrections have been neglected, we have not studied the DPAs for
the photon-induced channel.

\begin{table}
\begin{center} 
\begin{tabular}{ c  c  c  c  c}
 Ch. & $\sigma^{\rm WW \; DPA}_{\rm LO}$ [fb] & $\delta^{\rm WW \; DPA}_{\rm LO}$ [$ \% $] & $\sigma^{\rm tt \; DPA}_{\rm LO}$ [fb]  & $\delta^{\rm tt \; DPA}_{\rm LO}$ [$ \% $]\\
  \hline\hline
$\Pg \Pg$ &  $ 2.0003(1) $ & $ -0.56 $ & $ 1.9738(1) $ & $ -1.88 $  \\
${ q \bar{q}}$      &  $ 0.8437(5) $ & $ -0.58 $ & $ 0.83640(5) $ & $ -1.44 $ \\ 
  \hline
$\Pp\Pp$         &  $ 2.8441(1) $ & $ -0.56 $ & $ 2.8102(1) $ & $ -1.75 $ \\
  \hline
\end{tabular}
\end{center}
        \caption[Comparison of the integrated cross section DPA]{\label{table:LO_DPA_results_summary}
                Integrated LO cross sections for the two DPAs.
                The relative difference is defined as $\delta^{\rm DPA}_{\rm LO} = \sigma^{\rm DPA}_{\rm LO} / \sigma^{\rm Full}_{\rm LO} - 1 $ in per cent.}
\end{table}
 
For the NLO EW DPAs, only the two channels that have been computed in
the DPAs are shown in \refta{table:DPA_results_summary}.  Since the
${\rm g} q(/\bar{q})$ interference channels do not have virtual
corrections, we have not applied a DPA to them.  Interestingly, both
approximations reproduce the total cross section within a per mille.
Such an observation has already been made in
\citere{Denner:2016jyo} when computing the NLO EW corrections to
the off-shell production of two top quarks.  The reason is that the
Born, real-subtracted part and the convolution operator ($P$ and $K$
operator in \citere{Catani:1996vz}) contributions have been computed
with the full off-shell kinematics.  The DPA is only applied to the
virtual corrections and the $I$-operator which in our set-up amount to
$-0.2\%$ of the NLO predictions for the $\Pg \Pg$ channel.
 
\begin{table}
\begin{center} 
\begin{tabular}{ c  c  c  c  c}
 Ch. & $\sigma^{\rm WW \; DPA}_{\rm NLO \; EW}$ [fb] & $\delta^{\rm WW\; DPA}_{\rm NLO \; EW}$ [$ \% $] & $\sigma^{\rm tt \; DPA}_{\rm NLO \; EW}$ [fb]  & $\delta^{\rm tt \; DPA}_{\rm NLO \; EW}$ [$ \% $]\\
  \hline\hline
$\Pg \Pg$ &  $ 2.0237(6) $ & $ +0.18 $ & $ 2.0188(2) $ & $ -0.06 $  \\
${ q \bar{q}}$      &  $ 0.8470(4) $ & $ +0.19 $ & $ 0.8446(4) $ & $ -0.09 $ \\ 
  \hline
$\Pp\Pp$         &  $ 2.8712(8) $ & $ +0.18 $ & $ 2.8639(4) $ & $ -0.07 $ \\
  \hline
\end{tabular}
\end{center}
        \caption[Comparison of the integrated cross section DPA]{\label{table:DPA_results_summary}
                Integrated NLO cross section for the two DPAs.
                Only the channels where the DPAs are applied are shown.
                The relative difference is defined as $\delta^{\rm DPA}_{\rm NLO} = \sigma^{\rm DPA}_{\rm NLO} / \sigma^{\rm Full}_{\rm NLO} - 1 $ in per cent.}
\end{table}

\subsubsection*{Differential distributions}

\begin{figure}
        \setlength{\parskip}{-10pt}
        \begin{subfigure}{0.49\textwidth}
                \subcaption{}
                \includegraphics[width=\textwidth]{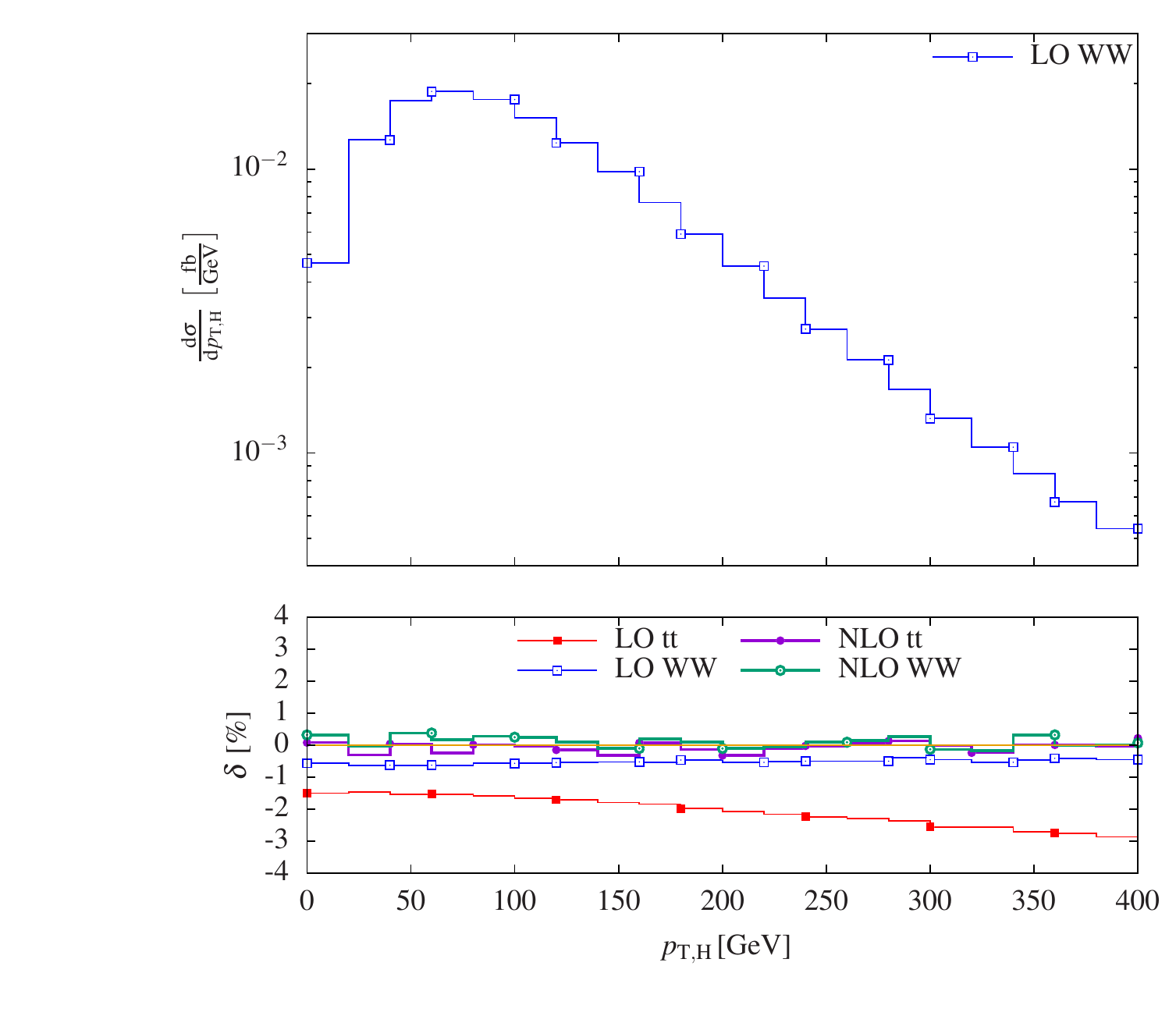}
                \label{plot:transverse_momentum_higgs_DPA}
        \end{subfigure}
        \hfill
        \begin{subfigure}{0.49\textwidth}
                \subcaption{}
                \includegraphics[width=\textwidth]{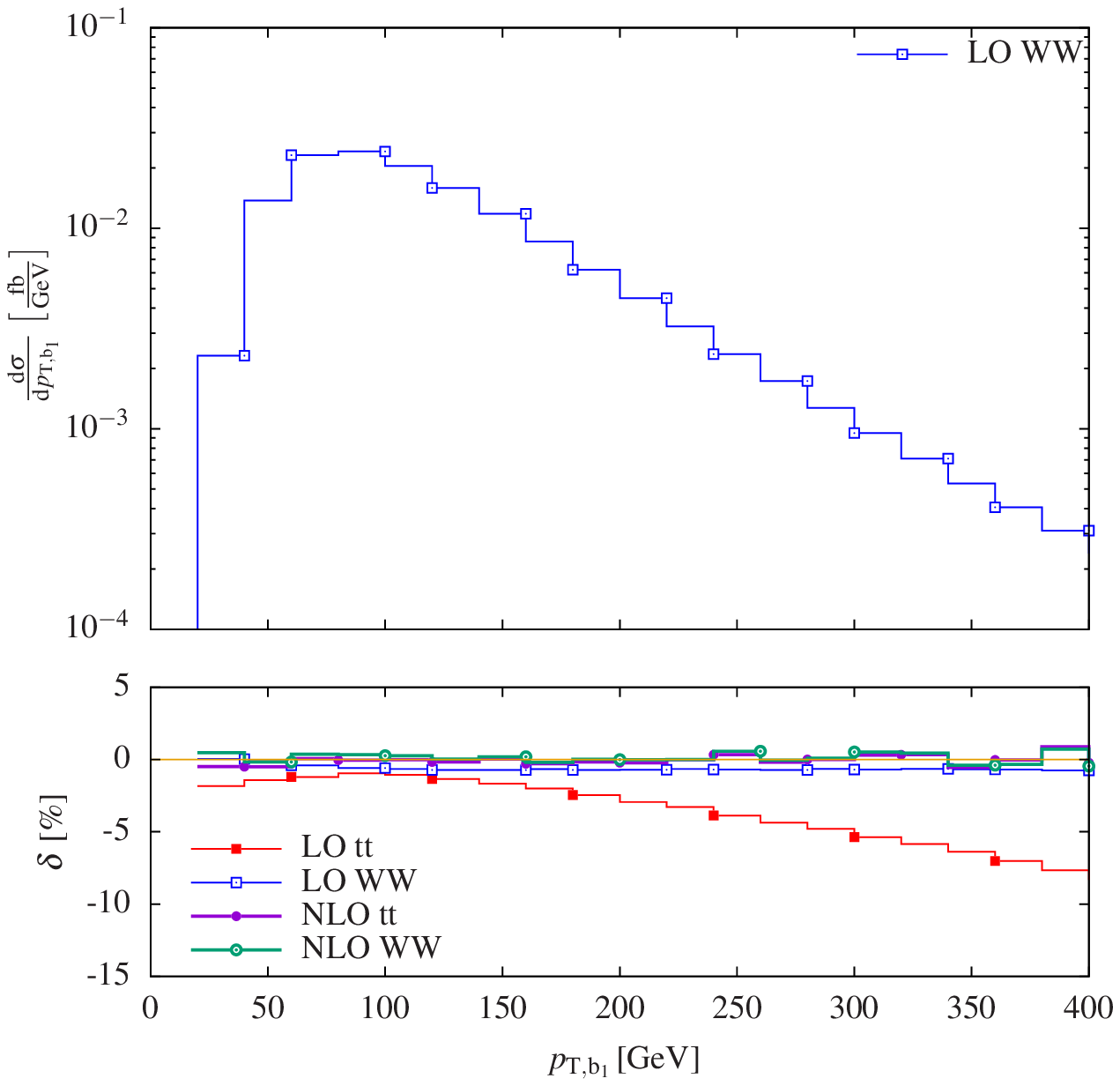}
                \label{plot:transverse_momentum_b1_DPA}
        \end{subfigure}

        \begin{subfigure}{0.49\textwidth}
                \subcaption{}
                \includegraphics[width=\textwidth]{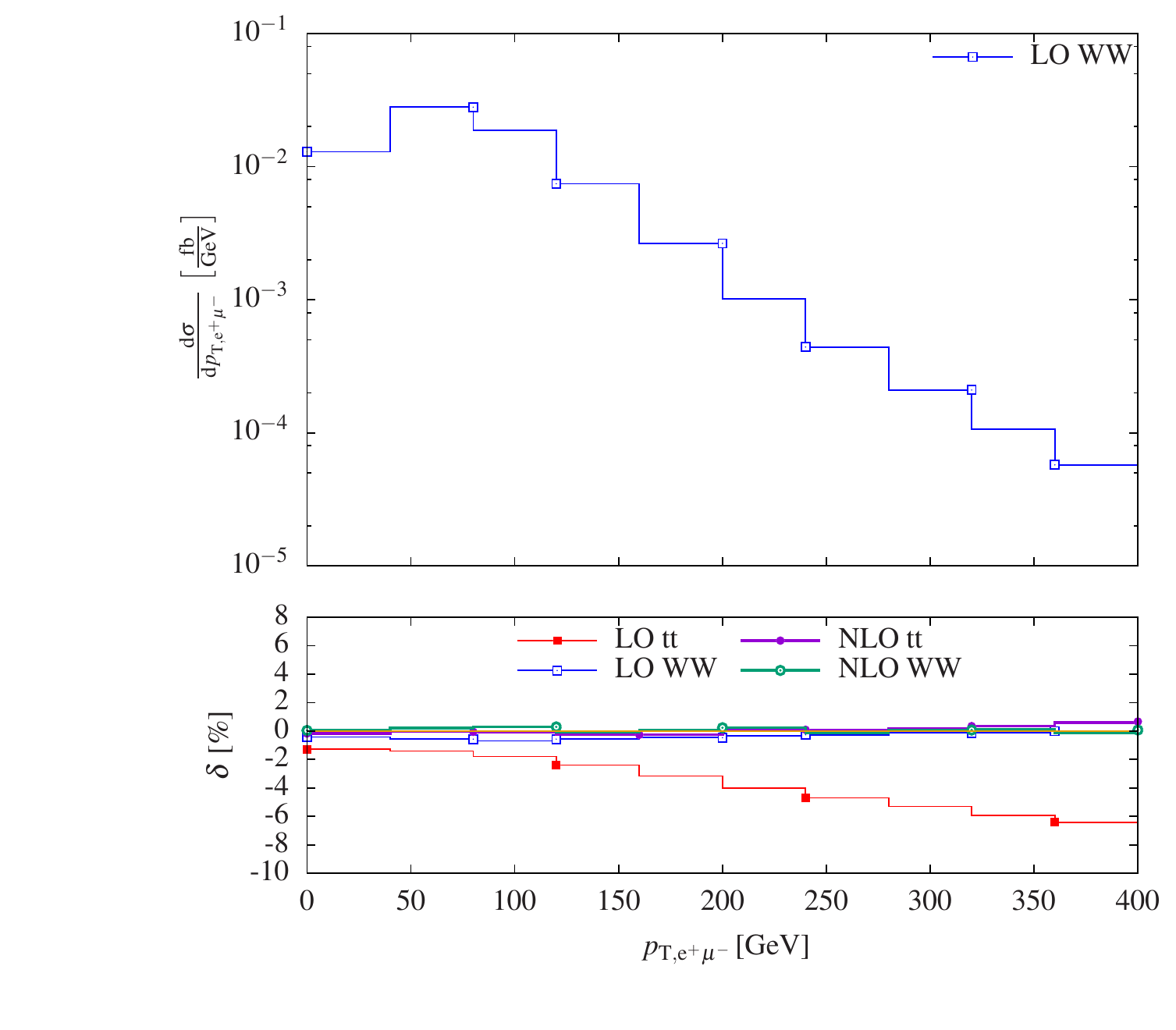}
                \label{plot:transverse_momentum_mupo_DPA}
        \end{subfigure}
        \hfill
        \begin{subfigure}{0.49\textwidth}
                \subcaption{}
                \includegraphics[width=\textwidth]{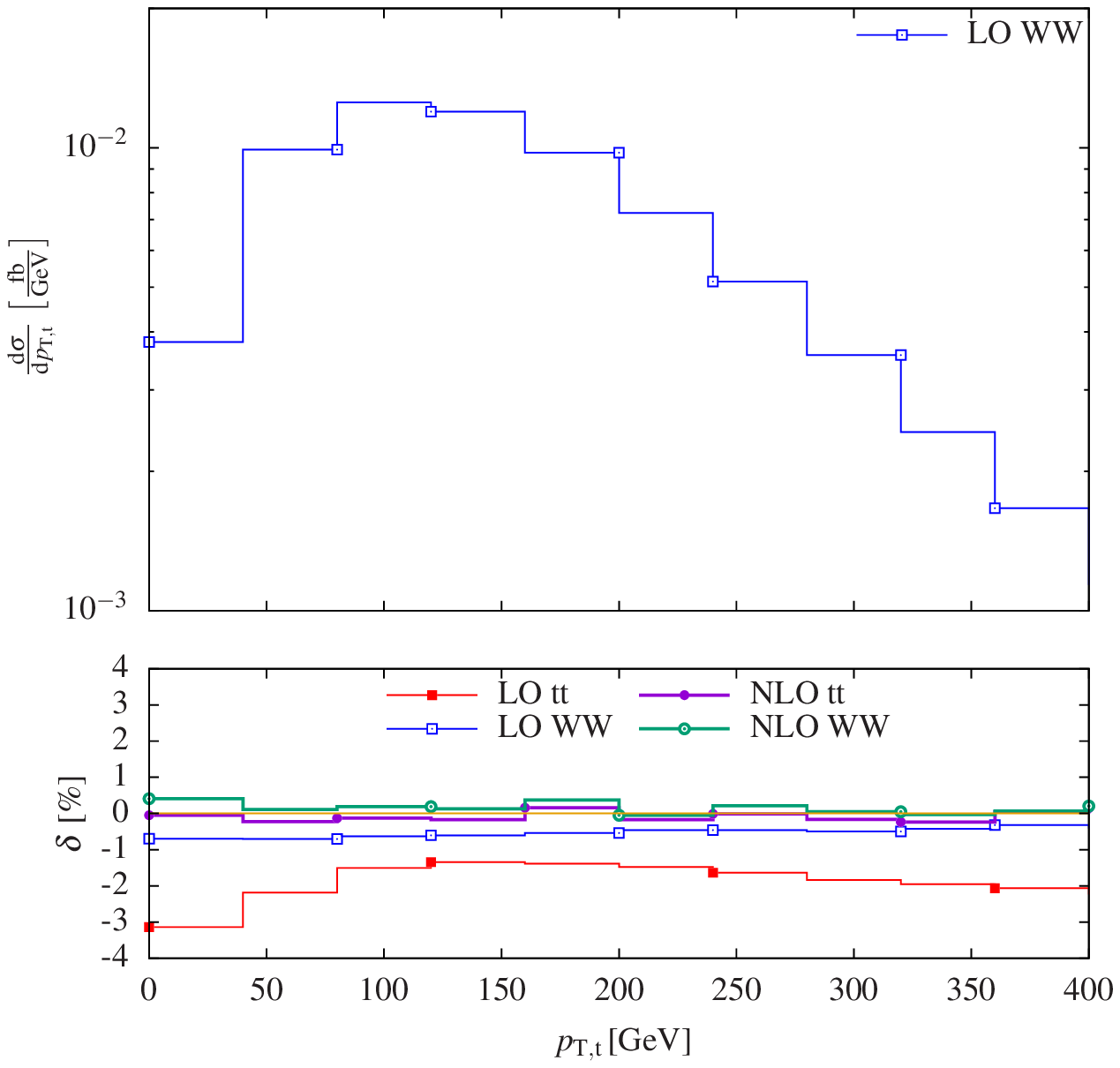}
                \label{plot:transverse_momentum_top_DPA} 
        \end{subfigure}

        \begin{subfigure}{0.49\textwidth}
                \subcaption{}
                \includegraphics[width=\textwidth]{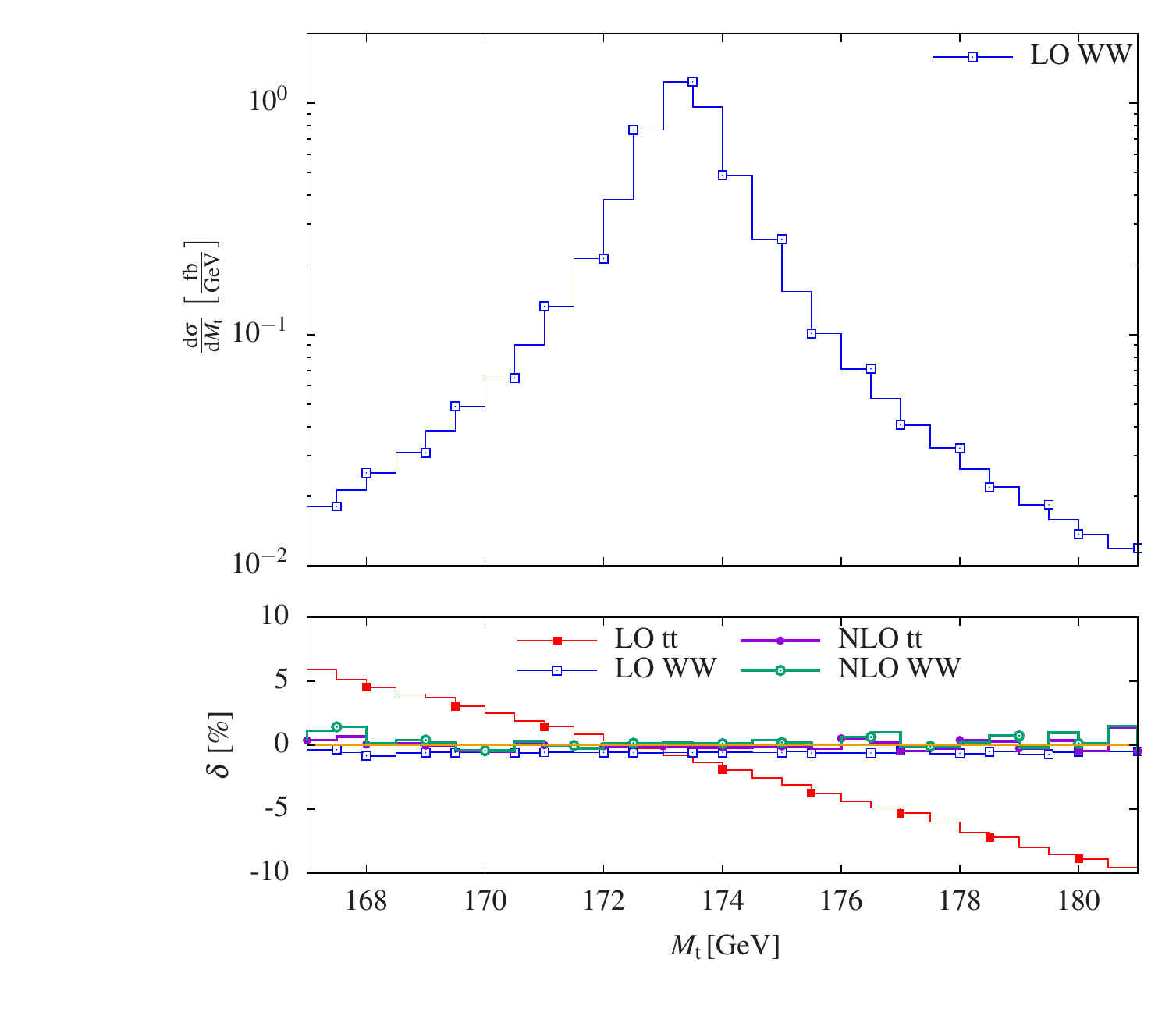}
                \label{plot:invariant_mass_truth_top_DPA}
        \end{subfigure}
        \hfill
        \begin{subfigure}{0.49\textwidth}
                \subcaption{}
                \includegraphics[width=\textwidth]{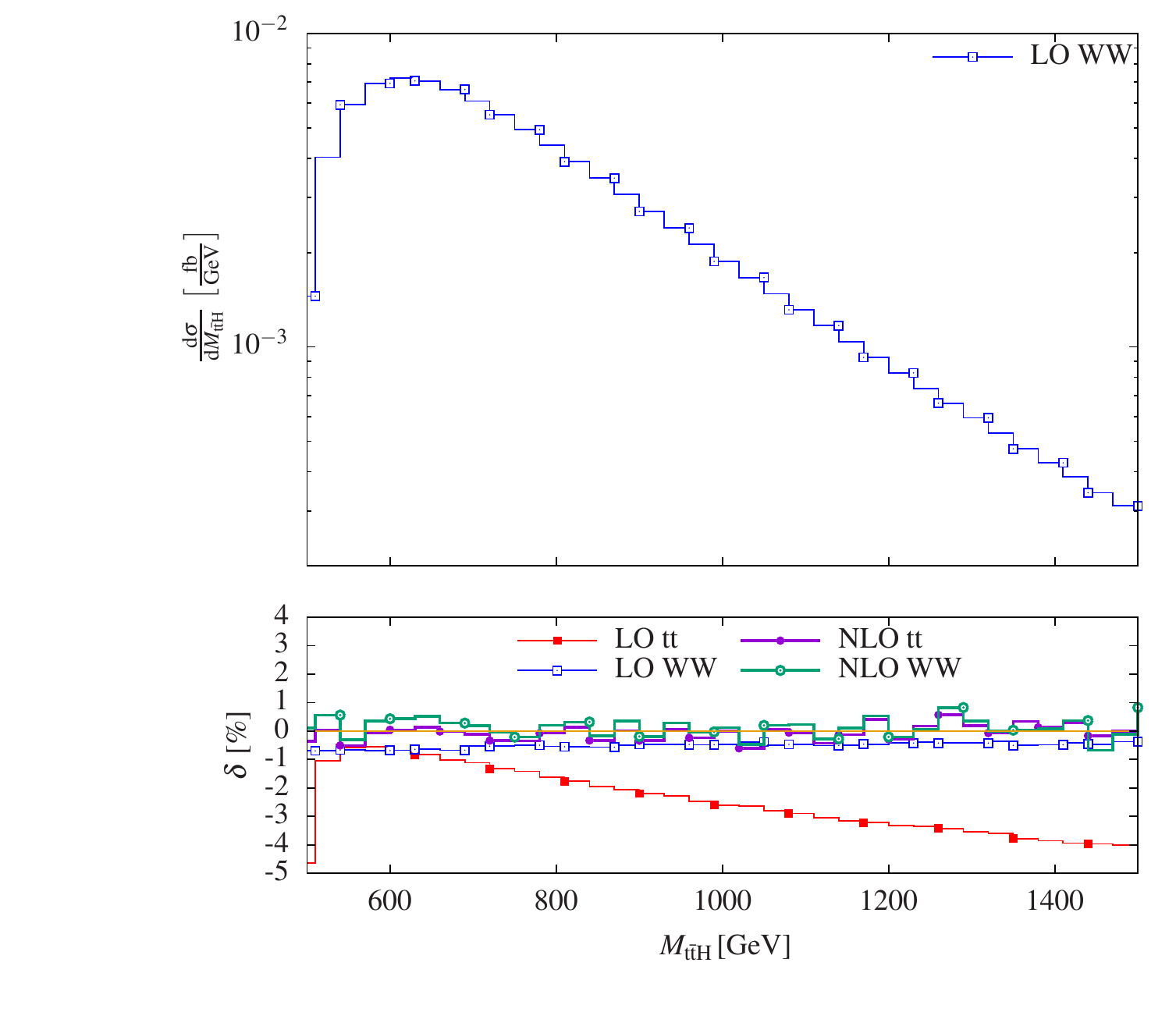}
                \label{plot:invariant_mass_truth_ttx_DPA}
        \end{subfigure}

        \vspace*{-3ex}
        \caption{\label{fig:transverse_momentum_distributions_DPA}%
          Comparison of the full calculation and the DPAs for various
          distributions at the LHC running at a centre-of-mass energy $\sqrt{s}=13\TeV$: 
          \subref{plot:transverse_momentum_higgs_DPA}~transverse momentum of the Higgs boson~(upper left), %
                \subref{plot:transverse_momentum_b1_DPA}~transverse momentum of the harder bottom jet~(upper right), %
                \subref{plot:transverse_momentum_mupo_DPA}~transverse momentum of the $\Pe^+ \mu^-$ system~(middle left) %
                \subref{plot:transverse_momentum_top_DPA}~transverse momentum  of reconstructed top quark~(middle right), %
                \subref{plot:invariant_mass_truth_top_DPA}~invariant mass of the reconstructed top quark~(lower left), and
                \subref{plot:invariant_mass_truth_ttx_DPA}~invariant mass of the reconstructed $\Pt\bar{\Pt}\PH$ system~(lower right).
                In the upper panels the LO distributions for the WW DPA are shown.
                The lower panels display the relative deviation of the different DPAs from the full
                calculation, $\delta = \sigma_{\text{DPA} } / \sigma_\text{Full} - 1$, in per cent.}
\end{figure}

In \reffi{fig:transverse_momentum_distributions_DPA}, the full
calculation is compared with the two DPAs at the distribution level at
both LO and NLO.  In the upper panels, only the WW DPA at LO is
displayed (as on a logarithmic scale the three other curves are
indistinguishable).  In the lower panels, the relative difference between
the approximations and the full computation at both LO and NLO is
shown.
The deviation with respect to the full calculation is defined
as $\delta = \sigma_{\text{DPA} } / \sigma_\text{Full} - 1$ and
expressed in per cent.
\change{Hence, in the lower panel ``LO tt'' denotes $\sigma^{\rm tt \; DPA}_{\rm LO} / \sigma^{\rm Full}_{\rm LO} - 1$, while ``NLO WW'' denotes $\sigma^{\rm WW \; DPA}_{\rm NLO} / \sigma^{\rm Full}_{\rm NLO} - 1$ for example.}
We recall that for the NLO prediction, the
DPA is not applied to the LO contributions, the real corrections, and
to the $P$- and $K$-operator terms.  The LO DPAs are, thus, not used in
the computation of the NLO DPAs and are shown here only for reference.

The transverse momentum distributions of the Higgs boson
(\reffi{plot:transverse_momentum_higgs_DPA}), of the harder bottom jet
(\reffi{plot:transverse_momentum_b1_DPA}), and of the $\Pe^+ \mu^-$
system (\reffi{plot:transverse_momentum_mupo_DPA}) exhibit similar
features at LO and NLO for both approximations.  The WW DPA
constitutes a better approximation than the tt one at LO and agrees
with the full calculation within $1 \%$ over the full range.
The LO tt DPA can deviate by up to $8\%$ for the harder bottom jet at $400\GeV$.
On the other hand, at NLO, both approximations reproduce the full
computation well.  As for the total cross section, the reason is that
the DPA is only applied to the subtracted virtual corrections.  In
addition, we only show distributions up to $400\GeV$ where off-shell
and non-resonant contributions are not sizeable yet.  But for higher
transverse momenta, as \emph{e.g.}~in \citere{Denner:2016jyo} for
off-shell top-antitop production, larger deviations from the full
calculation can be observed.
For example, for the NLO tt DPA, the difference can reach $4\%$ for the
transverse momentum of the harder bottom jet at $800\GeV$.  The
disagreement is less stringent here since the NLO EW corrections are
smaller and thus the relative difference is also smaller.

The distribution in the transverse momentum of the reconstructed top
quark is displayed in \reffi{plot:transverse_momentum_top_DPA}.  All
approximations agree within one per cent apart from the LO tt DPA.
The LO tt DPA is singled out from the other approximations
because of its normalisation that disagrees by $1.75\%$ with respect
to the full LO.  At zero transverse momentum, the disagreement is
about $3\%$ while it is $1.4\%$ around $150\GeV$ where the bulk of the
distribution is located.

The invariant-mass distribution of the reconstructed top quark shown
in \reffi{plot:invariant_mass_truth_top_DPA} is interesting.  All
approximations agree with the full computation within $\pm1\%$ apart
from the LO tt DPA.  At $167\GeV$, the LO tt DPA exceeds the full LO
prediction by $5\%$ while at $181\GeV$ it is lower by $10\%$.
The full computation at LO and the LO tt DPA agree
exactly at the top-quark mass value ($173.34\GeV$).

The invariant mass of the reconstructed $\Pt\bar{\Pt}\PH$ system
depicted in \reffi{plot:invariant_mass_truth_ttx_DPA} shows similar
features than the transverse-momentum distributions.  The LO tt DPA
tends to diverge from the full calculation towards higher invariant
masses to disagree by almost $5\%$ at $1500\GeV$.  The LO WW DPA on
the other hand agrees perfectly over the full range.  At NLO, both
approximations describe the full computation well even at an invariant
mass of $1500\GeV$.

No rapidity distributions are displayed here as none of them shows any
shape distortion between neither of the two DPAs and the full
calculation.  Differences for the LO tt DPA are only due to the
different normalisation.  For the distributions in the azimuthal-angle
separation and the cosine of the angle between the two leptons, the
shape distortions are also below one per cent.

To conclude, the tt DPA does not always provide a good description
of the full calculation.  This is particularly apparent in phase-space
regions where off-shell and non-resonant contributions are sizeable.
In these region, the DPA requiring two resonant top quarks can
disagree by up to $4\%$.
On the contrary, for all inspected distributions, the WW DPA describes
the full calculation within a per cent over the considered phase-space
range.

\subsection{Combination of NLO EW and QCD corrections}
\label{sec:CombinationNLO}

In this section, we present predictions for the integrated cross
section as well as distributions taking into account both NLO EW and
QCD corrections.  These can be considered as state-of-the-art
predictions for the production of a Higgs boson in association with a
pair of off-shell top quarks.  The NLO EW and QCD \change{cross sections} are
defined as:
\begin{equation}
\label{sigmadef}
 \sigma^{\mathrm{NLO}}_{\mathrm{QCD}} = \sigma^{\mathrm{Born}} + \delta \sigma^{\mathrm{NLO}}_{\mathrm{QCD}} \qquad \text{and} \qquad
 \sigma^{\mathrm{NLO}}_{\mathrm{EW}} = \sigma^{\mathrm{Born}} + \delta \sigma^{\mathrm{NLO}}_{\mathrm{EW}},
\end{equation}
\change{where the top width used in the top propagators includes both
  NLO QCD and EW corrections both at Born and NLO level.  This allows
  a straight-forward additive combination of the two types of
  corrections:}
\begin{equation}
\label{additionNLO}
 \sigma^{\mathrm{NLO}}_{\mathrm{QCD+EW}} = \sigma^{\mathrm{Born}} + \delta \sigma^{\mathrm{NLO}}_{\mathrm{QCD}} + \delta \sigma^{\mathrm{NLO}}_{\mathrm{EW}}.
\end{equation}
\change{Based on the definition of Eq.~(\ref{sigmadef}), the
  multiplicative combination can be defined as}
\begin{equation}
\label{productNLO}
 \sigma^{\mathrm{NLO}}_{\mathrm{QCD}\times\mathrm{EW}} = \sigma^{\mathrm{NLO}}_{\mathrm{QCD}} \left( 1 + \frac{\delta \sigma^{\mathrm{NLO}}_{\mathrm{EW}}}{\sigma^{\mathrm{Born}}} \right)
 = \sigma^{\mathrm{NLO}}_{\mathrm{EW}} \left( 1 + \frac{\delta \sigma^{\mathrm{NLO}}_{\mathrm{QCD}}}{\sigma^{\mathrm{Born}}} \right) .
\end{equation}
The difference between these two ways of combining NLO EW and QCD corrections can give an estimate of the missing higher orders to the QCD--EW mixed contributions.
The NLO QCD$\times$EW combination can be understood as an improved prediction when the typical scales of the QCD and EW corrections are well separated.
This is, for example, the case for soft QCD interactions with a scale well below the EW scale.

In Table~\ref{table:combination_results_summary}, a summary of the NLO
QCD and EW corrections is presented.  Note that in the present
calculation $\sigma^{\mathrm{Born}}$ and $\sigma^{\mathrm{LO}}$ are
not identical.
\change{While $\sigma^{\mathrm{LO}}$ is computed with a top width without NLO corrections, $\sigma^{\mathrm{Born}}$ is calculated with a top width featuring both NLO EW and QCD corrections.}
The latter Born contribution enters the
cross sections $\sigma^{\mathrm{NLO}}_{\mathrm{QCD}}$ and
$\sigma^{\mathrm{NLO}}_{\mathrm{EW}}$ at NLO QCD and EW, respectively.
In addition the two possible ways of combining the NLO
corrections are also given.  The difference is negligible (of the
order of the Monte Carlo error) and thus indicates that the missing
higher orders to the QCD--EW mixed contributions are small.

\begin{table}
\begin{center} 
\begin{tabular}{ c  c  c  c  c  c}
 $\sigma^{\rm LO}$ & $\sigma^{\rm Born}$ & $\sigma^{\mathrm{NLO}}_{\mathrm{QCD}}$ & $\sigma^{\mathrm{NLO}}_{\mathrm{EW}}$ & $\sigma^{\mathrm{NLO}}_{\mathrm{QCD+EW}}$  & $\sigma^{\mathrm{NLO}}_{\mathrm{QCD}\times\mathrm{EW}}$ \\
  \hline\hline
$ 2.4817(1) $ &  $ 2.7815(1) $ & $ 2.866(1) $ & $ 2.721(3) $ & $ 2.806 $ & $ 2.804 $ \\
  \hline
\end{tabular}
\end{center}
        \caption[Summary of the integrated cross section]{\label{table:combination_results_summary}
                Integrated cross sections for $\Pp\Pp \to \Pe^+\nu_\Pe \mu^- \bar{\nu}_\mu \Pb \bar{\Pb} \PH (\gamma / \Pj)$ at the LHC at a centre-of-mass energy of $\sqrt{s}=13\TeV$.
                For $\sigma^{\rm LO}$, no NLO corrections are included in the top-quark width while for $\sigma^{\rm Born}$ both NLO EW and QCD corrections are included.
                The latter Born contribution is the one used when computing the QCD and EW NLO predictions $\sigma^{\mathrm{NLO}}_{\mathrm{QCD}}$ and $\sigma^{\mathrm{NLO}}_{\mathrm{EW}}$, respectively.
                In addition, two ways of combining the NLO corrections are presented.
                They are defined in Eqs.~\eqref{additionNLO} and \eqref{productNLO} for QCD+EW and QCD$\times$EW, respectively.
                All cross sections are expressed in femtobarn (fb).}
\end{table}

We do not discuss the effects of the NLO QCD corrections as they have
been investigated in detail in \citere{Denner:2015yca}, in
particular, scale uncertainty and various differential distributions.
We only discuss here results for the combination of both NLO EW and
QCD corrections.  Hence, in the upper panels the LO, NLO QCD+EW as
well as the NLO QCD$\times$EW predictions are displayed.  In addition
to predictions for the central scale we have also calculated
predictions where both the renormalisation and factorisation scale are
scaled by a factor 2 up and down.
The envelope is obtained by taking the minimum and the maximum of
these three predictions.  In the lower panels the ratio of the LO and
NLO predictions with respect to the LO prediction at the central scale
is shown. Thus, the central curve for NLO corresponds to the usual $K$
factor $\mathrm{K} = \sigma_\text{NLO} / \sigma_\text{LO}$. The bands
in the lower panels show the scale uncertainty resulting from the LO
or NLO cross section in the numerator.

\begin{figure}
        \setlength{\parskip}{-10pt}
        
        \begin{subfigure}{0.49\textwidth}
                \subcaption{}
                \includegraphics[width=\textwidth]{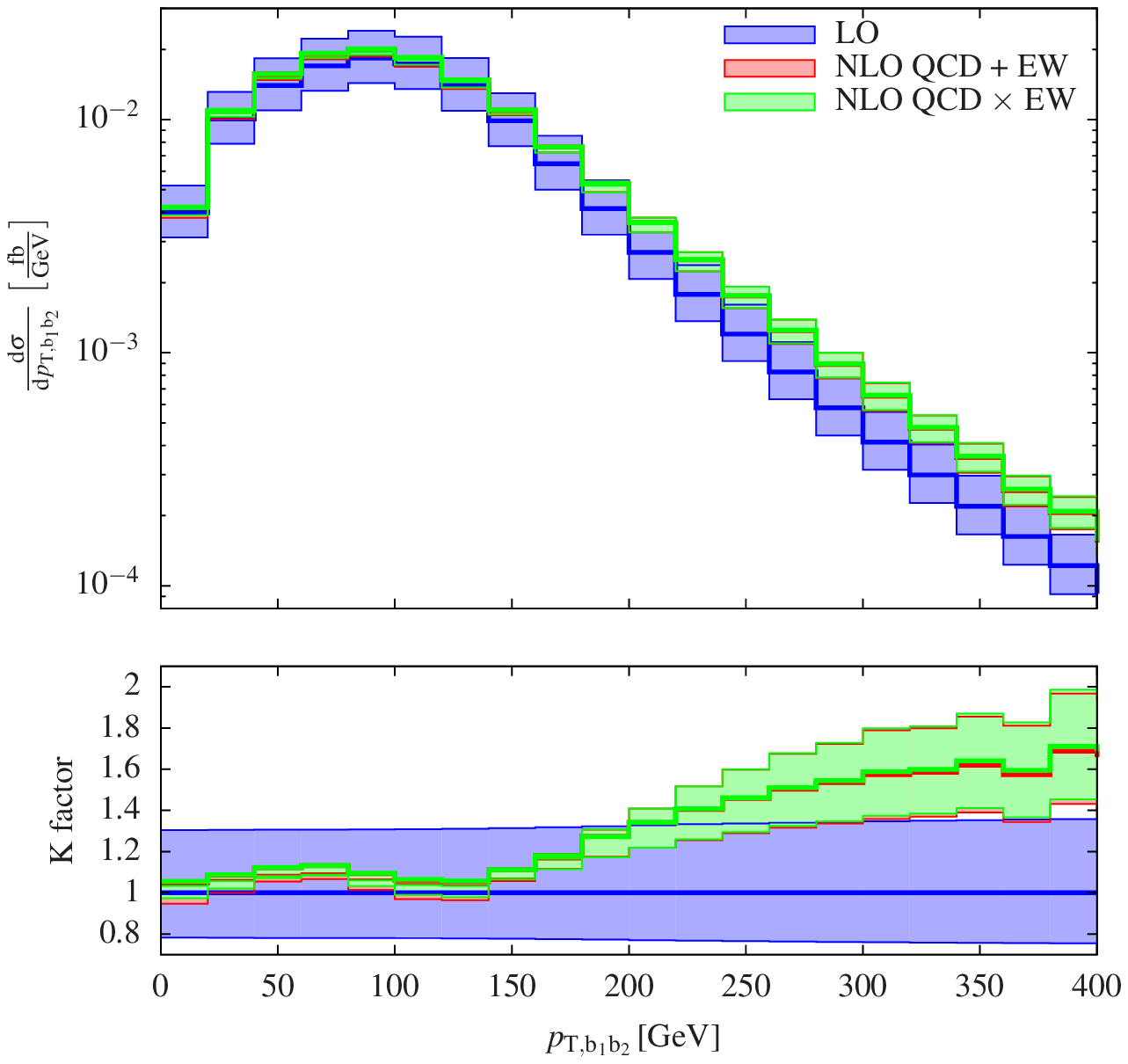}
                \label{plot:transverse_momentum_bb12_qcdew}
        \end{subfigure}        
        \hfill
        \begin{subfigure}{0.49\textwidth}
                \subcaption{}
                \includegraphics[width=\textwidth]{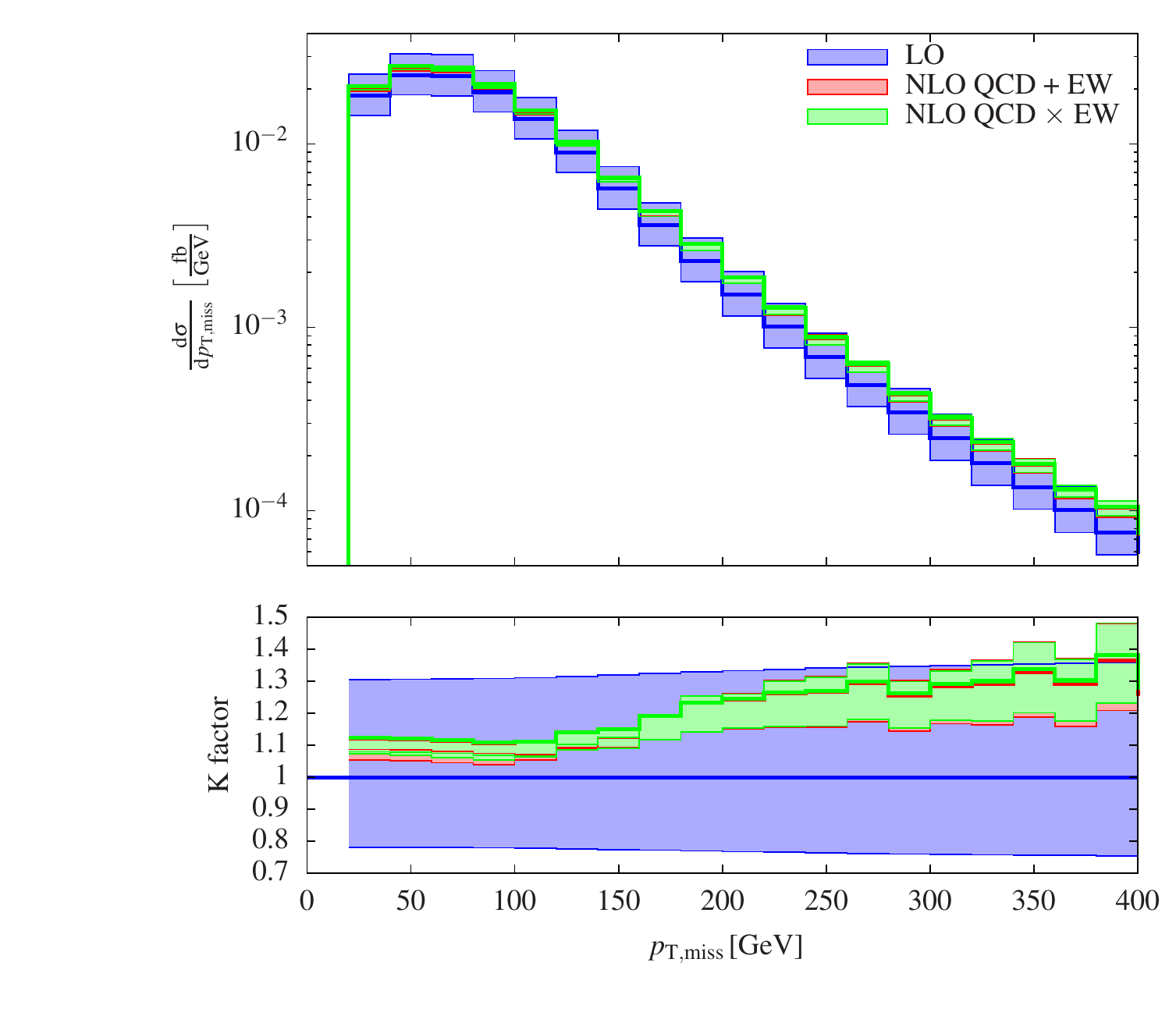}
                \label{plot:transverse_momentum_truth_missing_qcdew} 
        \end{subfigure}

        \begin{subfigure}{0.49\textwidth}
                \subcaption{}
                \includegraphics[width=\textwidth]{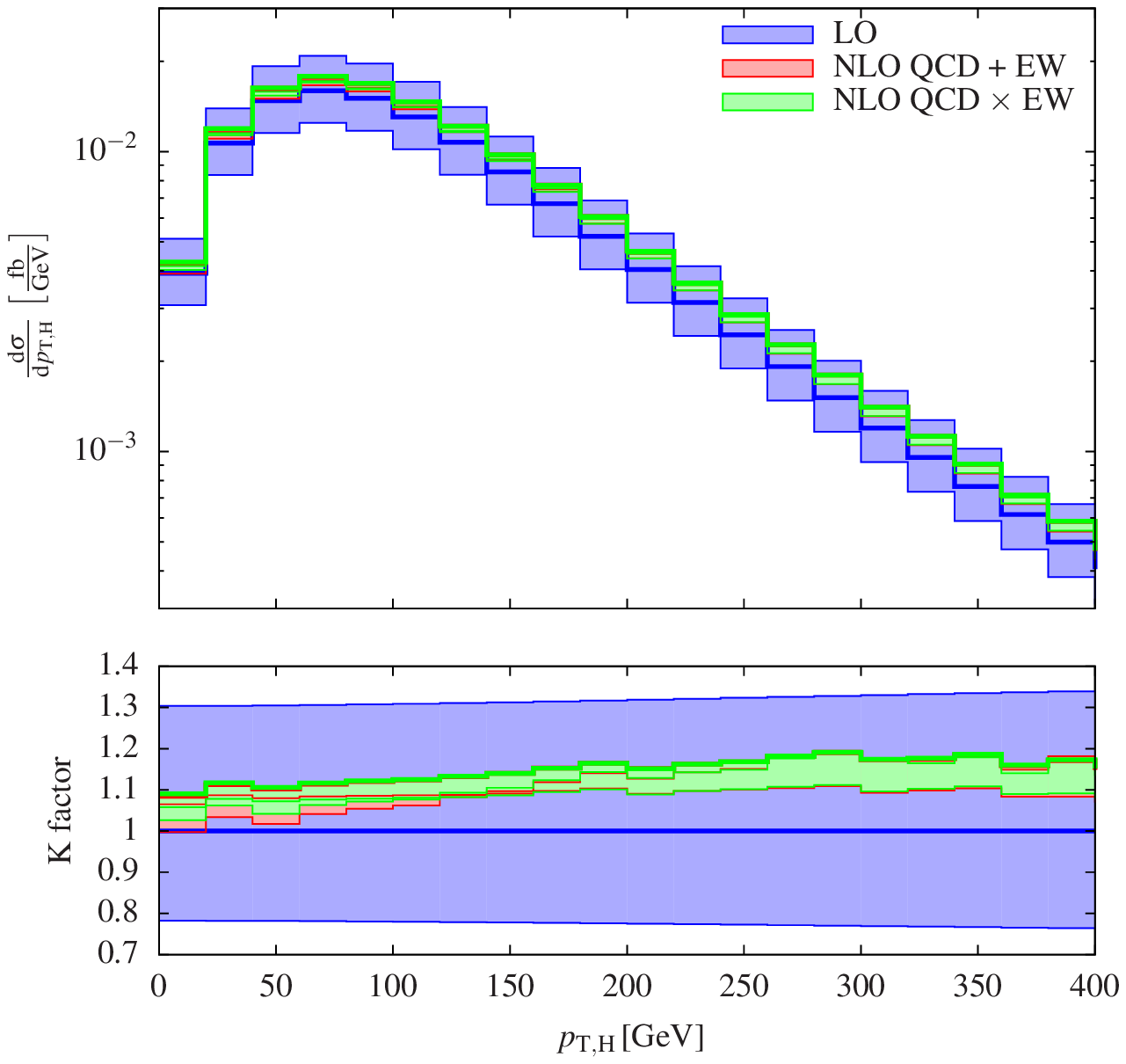}
                \label{plot:transverse_momentum_higgs_qcdew}
        \end{subfigure}
        \hfill
        \begin{subfigure}{0.49\textwidth}
                \subcaption{}
                \includegraphics[width=\textwidth]{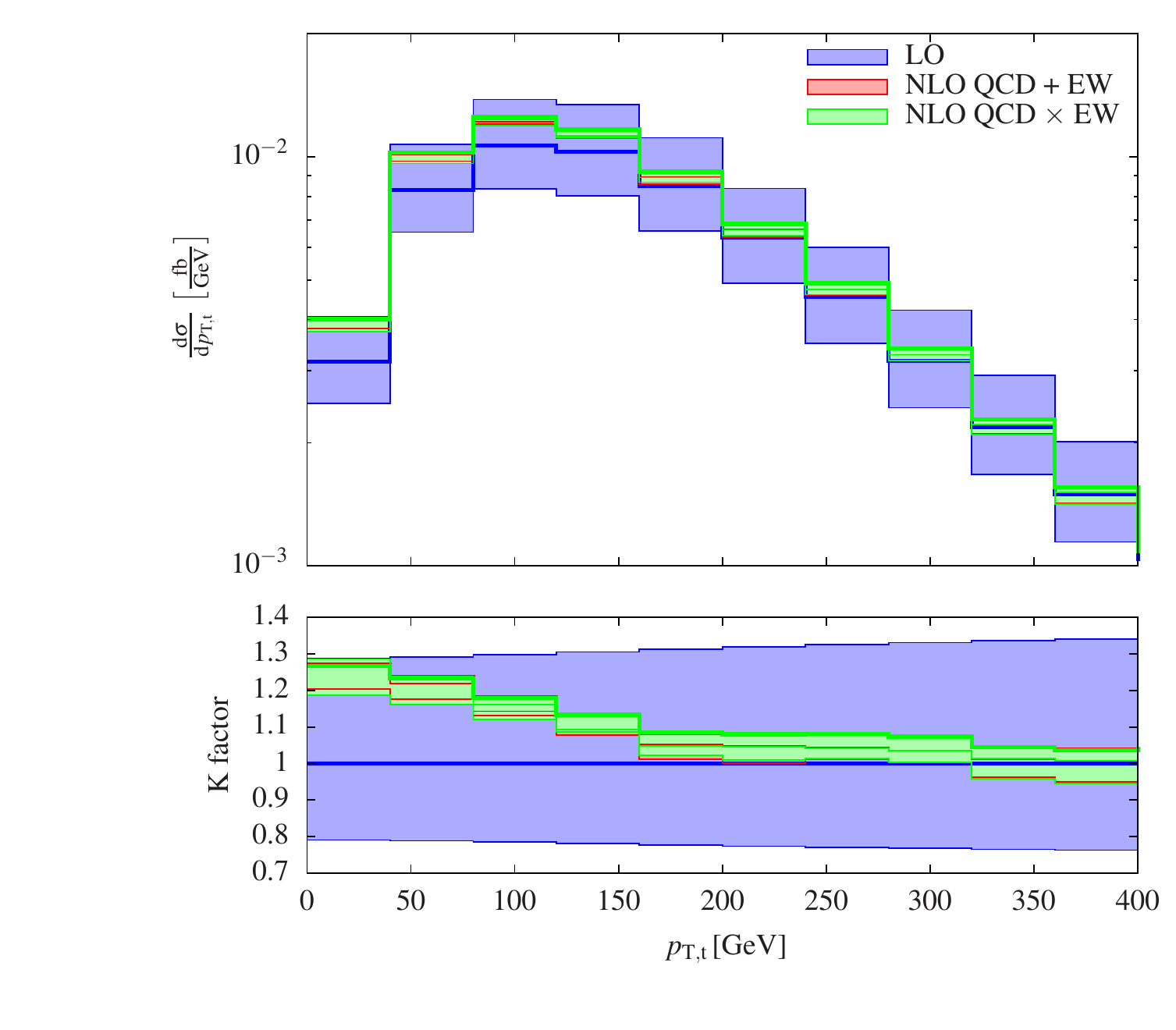}
                \label{plot:transverse_momentum_truth_top_qcdew}
        \end{subfigure}
        
        \begin{subfigure}{0.49\textwidth}
                \subcaption{}
                \includegraphics[width=\textwidth]{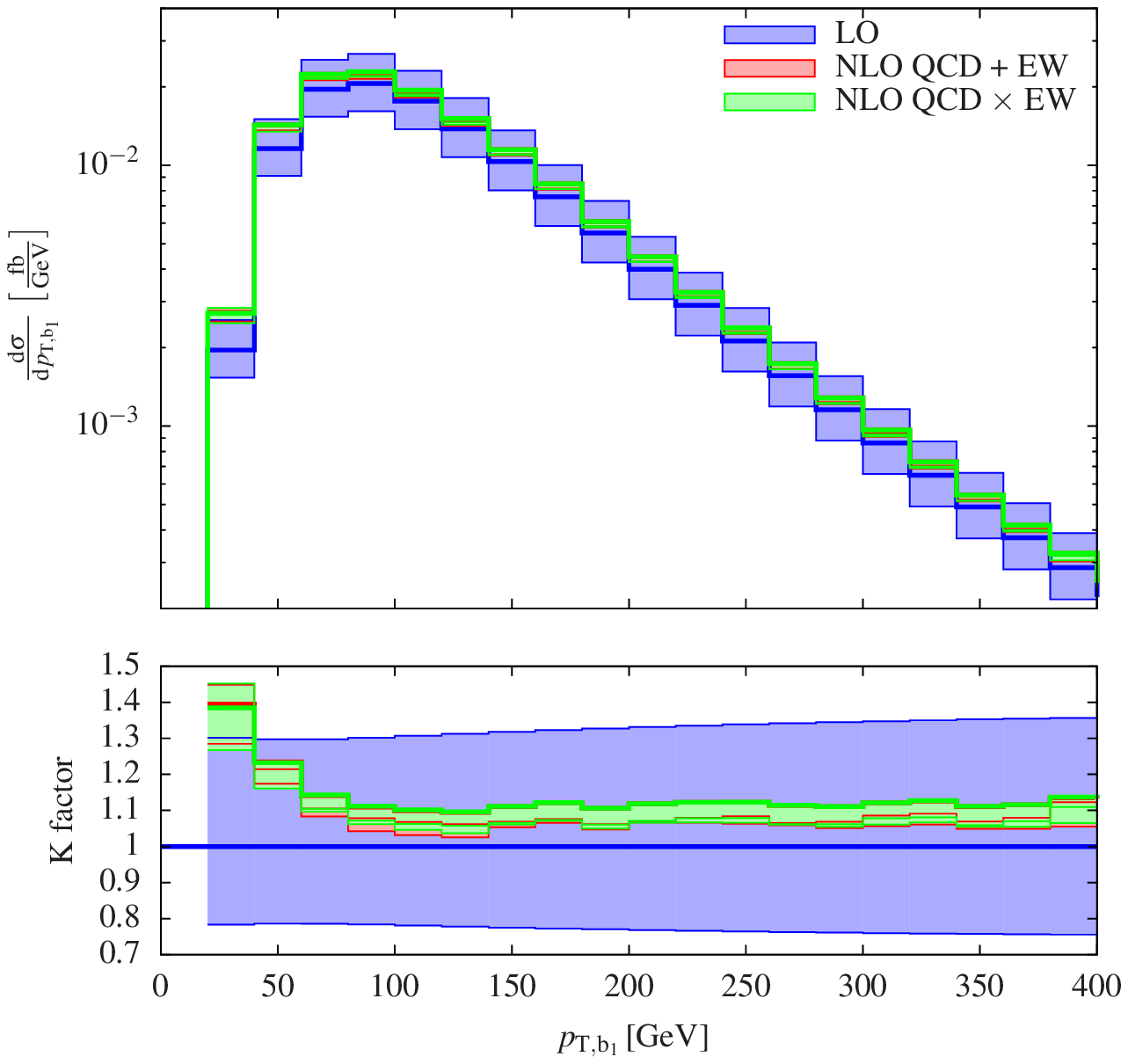}
                \label{plot:transverse_momentum_b1_qcdew}
        \end{subfigure}
        \hfill
        \begin{subfigure}{0.49\textwidth}
                \subcaption{}
                \includegraphics[width=\textwidth]{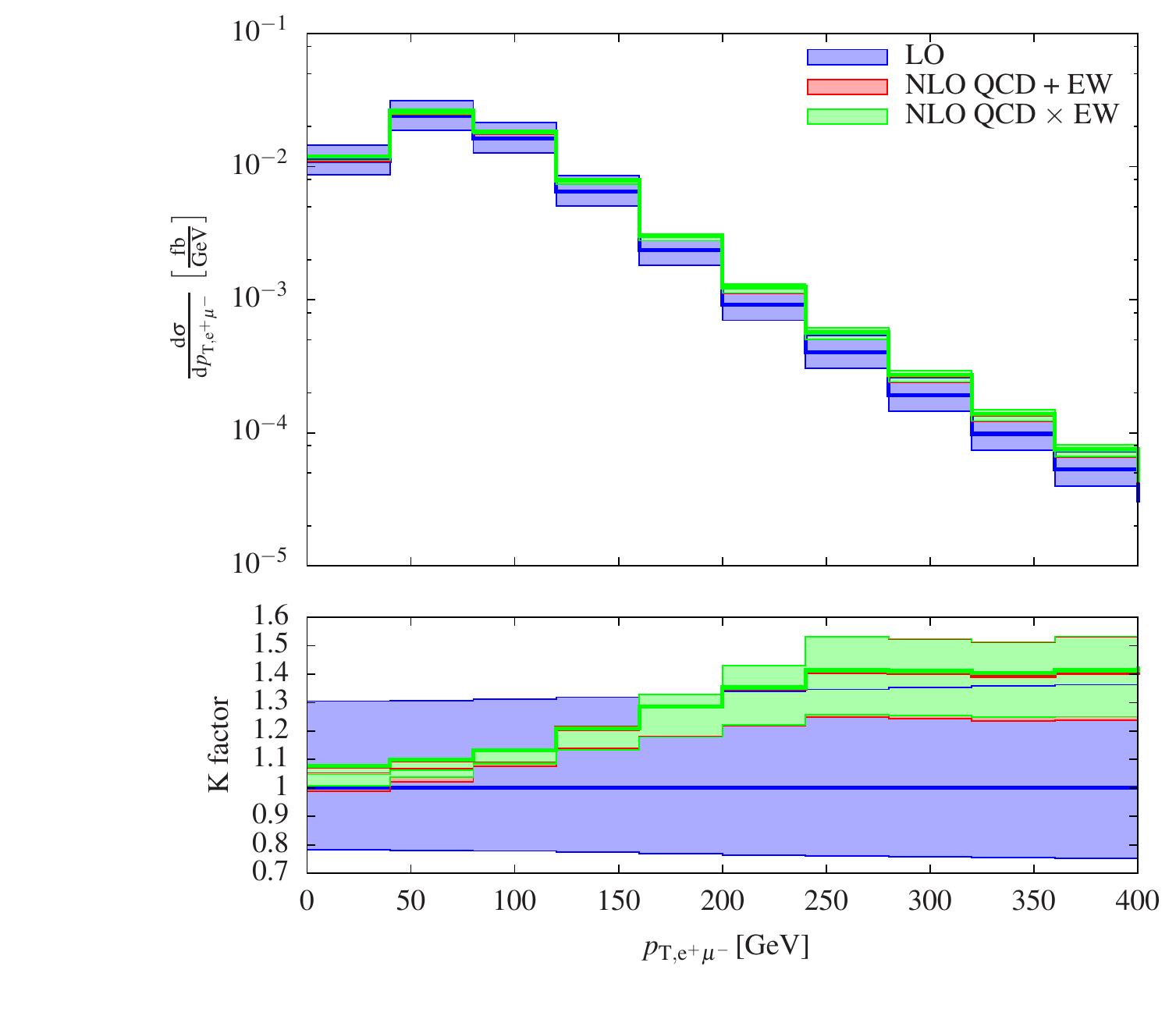}
                \label{plot:transverse_momentum_truth_mupo_qcdew}
        \end{subfigure}
     
        \vspace*{-3ex}
        \caption{\label{fig:transverse_momentum_distributions_combined}%
          Combined NLO EW and QCD corrections for various
          distributions at the LHC running at a centre-of-mass energy $\sqrt{s}=13\TeV$: 
          \subref{plot:transverse_momentum_bb12_qcdew} transverse momentum for the bottom-jet pair~(upper left), %
          \subref{plot:transverse_momentum_truth_missing_qcdew} for the missing momentum~(upper right), %
          \subref{plot:transverse_momentum_higgs_qcdew} for the Higgs boson~(middle left) %
          \subref{plot:transverse_momentum_truth_top_qcdew} for the reconstructed top quark~(middle right)
          \subref{plot:transverse_momentum_b1_qcdew} for the harder bottom~jet~(lower left), and %
          \subref{plot:transverse_momentum_truth_mupo_qcdew} for the ${\Pe}^+ \mu^-$ system~(lower right).
          In the upper panels the LO, the NLO QCD$+$EW and NLO QCD$\times$EW distributions are shown.
          The lower panels display the two differently combined NLO predictions with respect to the LO one as $\mathrm{K} = \sigma_\text{NLO} / \sigma_\text{LO}$.}
\end{figure}

In general, the NLO effects are mainly driven by the QCD corrections
as the EW corrections are smaller.  For instance, this can be observed
in the distribution of the bottom-jet pair shown
in~\reffi{plot:transverse_momentum_bb12_qcdew}.  At $400\GeV$ the
combined EW and QCD corrections amount to about $74\%$.  There, the EW
corrections reach $-6\%$ while the QCD ones reach almost $80\%$
\cite{Denner:2015yca}.  Thus, even if the EW
corrections are non-negligible, they look small by comparison to large
QCD corrections.  For this particular distribution, this huge effect
is due to a kinematical constraint at LO which forces the transverse
momenta of the bottom-jet--pair system to be strongly suppressed above
$150\GeV$.  At NLO, this kinematical constraint is relaxed by the
emission of an extra parton and thus leads to a huge K factor in this
phase-space region.

The difference between the NLO combinations QCD+EW and QCD$\times$EW
gives an estimate of the missing higher orders of QCD--EW mixed
contributions.  In this regard, the two different ways of combining
NLO effects give rather similar results.
Consider for example the distribution in the missing transverse energy
in \reffi{plot:transverse_momentum_truth_missing_qcdew} where the EW
corrections reach $8\%$.  At this very point, the two combinations
differ by roughly two per cent at $400\GeV$ indicating that missing
higher orders are expected to be small in this phase-space region.

Figures~\ref{plot:transverse_momentum_higgs_qcdew} and \ref{plot:transverse_momentum_truth_top_qcdew} depict the transverse-momentum distribution of the Higgs boson and of the reconstructed top quark, respectively.
These are key observables for a measurement of the Higgs production in association with top quarks.

Finally, the transverse momentum of the harder bottom jet and of the
${\Pe}^+ \mu^-$ system are shown in
Figures~\ref{plot:transverse_momentum_b1_qcdew} and
\ref{plot:transverse_momentum_truth_mupo_qcdew}, respectively.  As the
QCD corrections are positive, adding the EW corrections has the effect
of damping the QCD corrections in the high-energy regime where Sudakov
logarithms arise.

\begin{figure}
        \setlength{\parskip}{-10pt}
        \begin{subfigure}{0.49\textwidth}
                \subcaption{}
                \includegraphics[width=\textwidth]{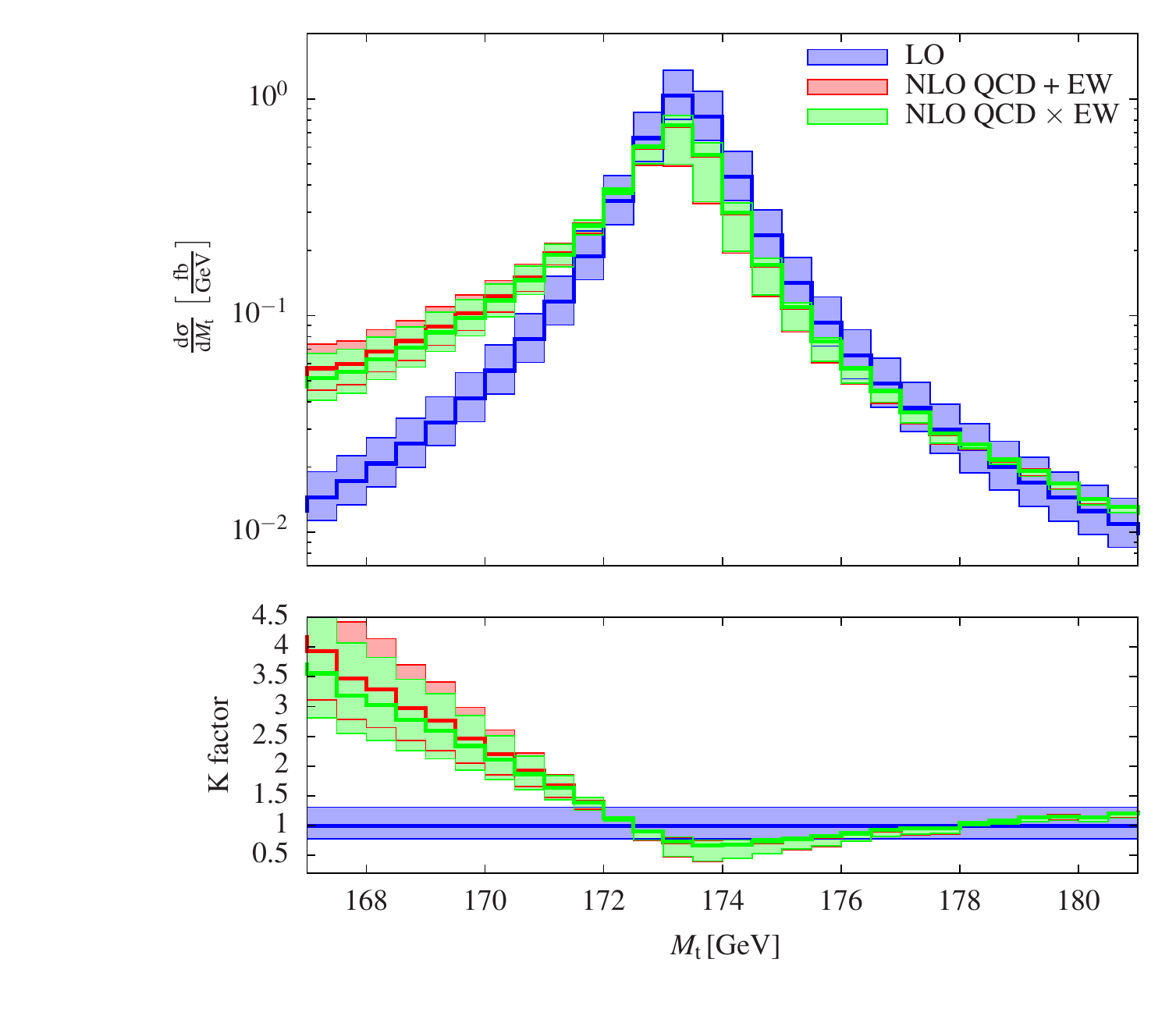}
                \label{plot:invariant_mass_truth_top_qcdew}
        \end{subfigure}
        \hfill
        \begin{subfigure}{0.49\textwidth}
                \subcaption{}
                \includegraphics[width=\textwidth]{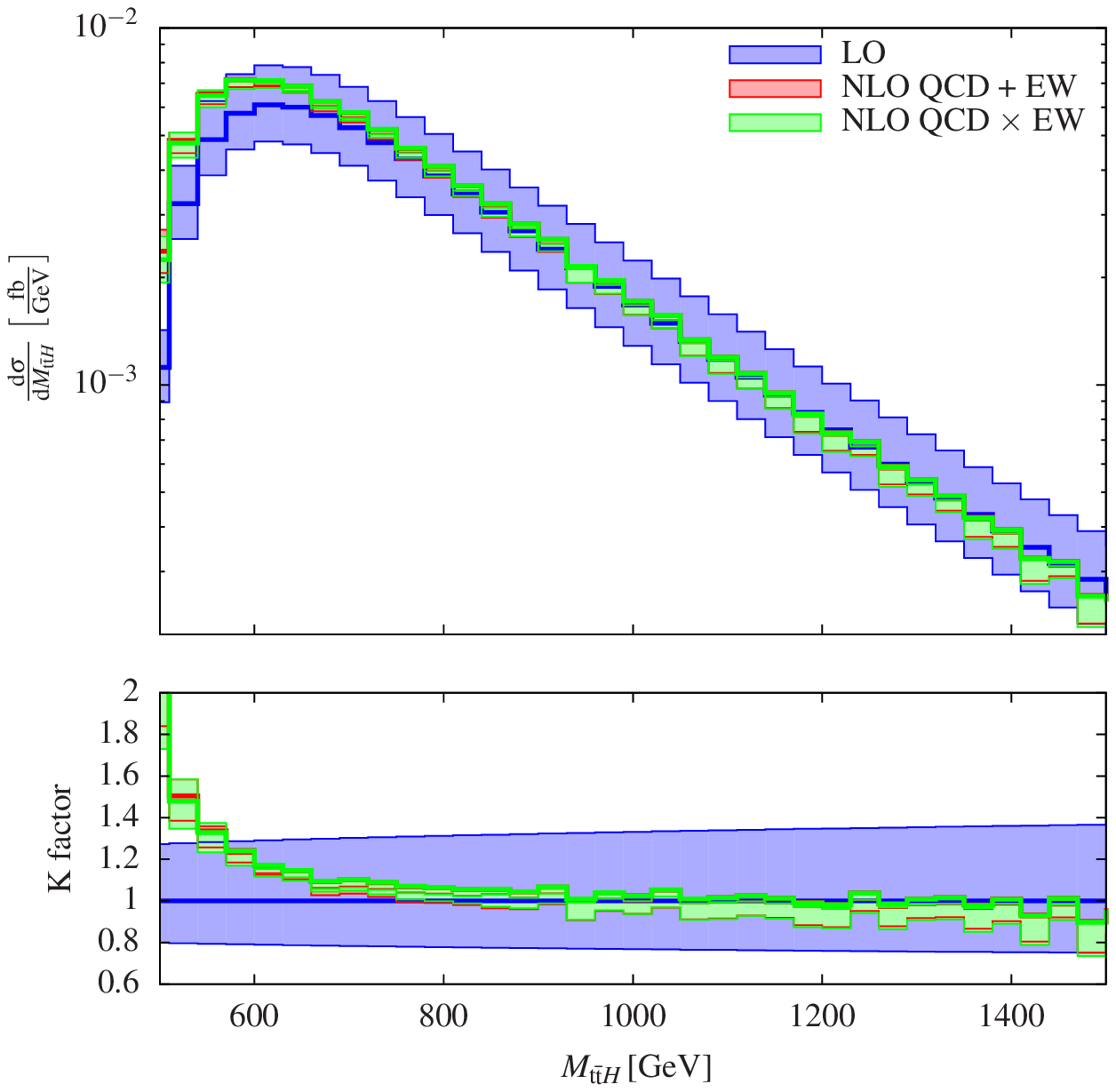}
                \label{plot:invariant_mass_truth_ttxh_qcdew}
        \end{subfigure}

        \begin{subfigure}{0.49\textwidth}
                \subcaption{}
                \includegraphics[width=\textwidth]{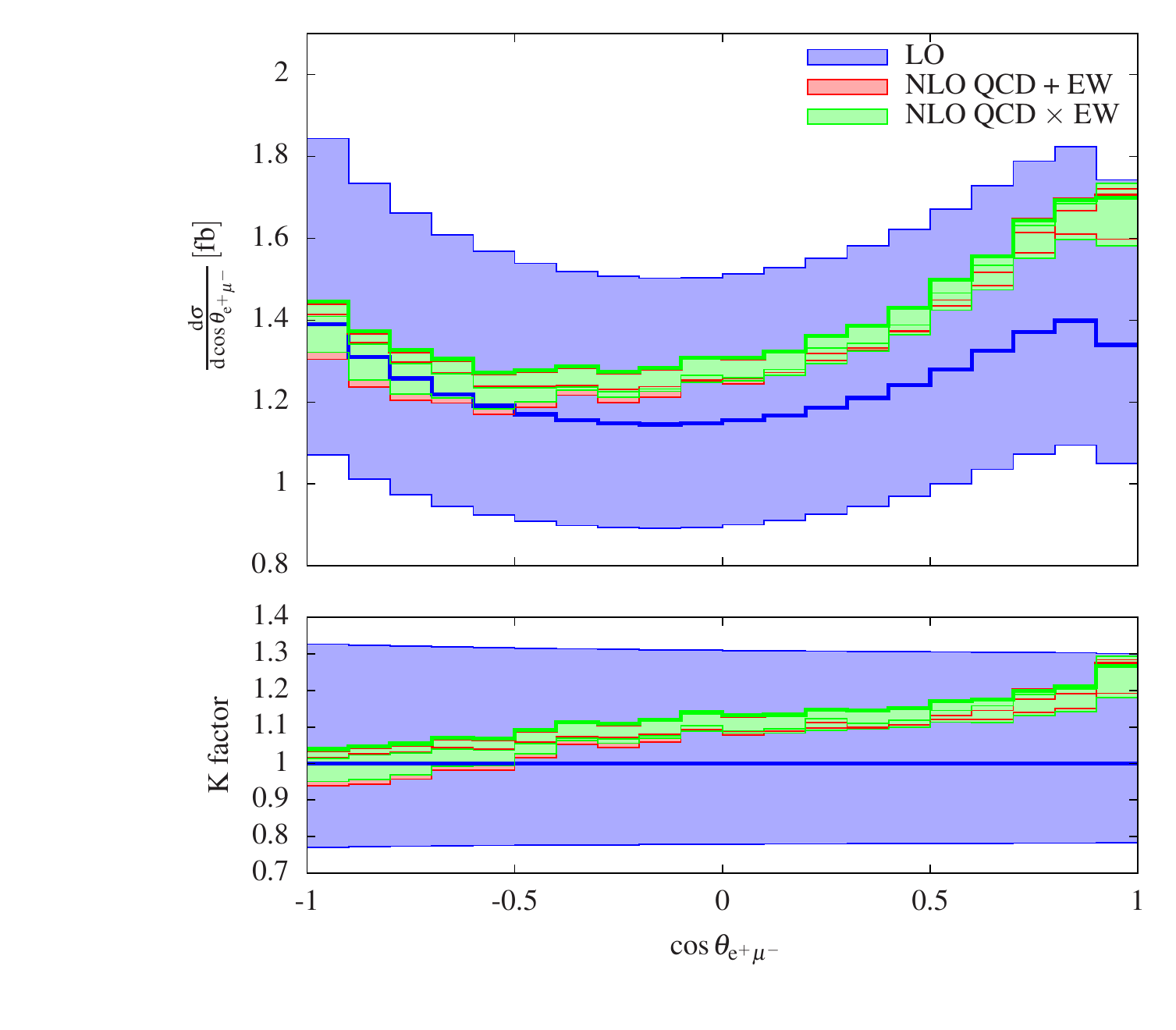}
                \label{plot:cosine_angle_separation_epmu_qcdew}
        \end{subfigure}
        \hfill
        \begin{subfigure}{0.49\textwidth}
                \subcaption{}
                \includegraphics[width=\textwidth]{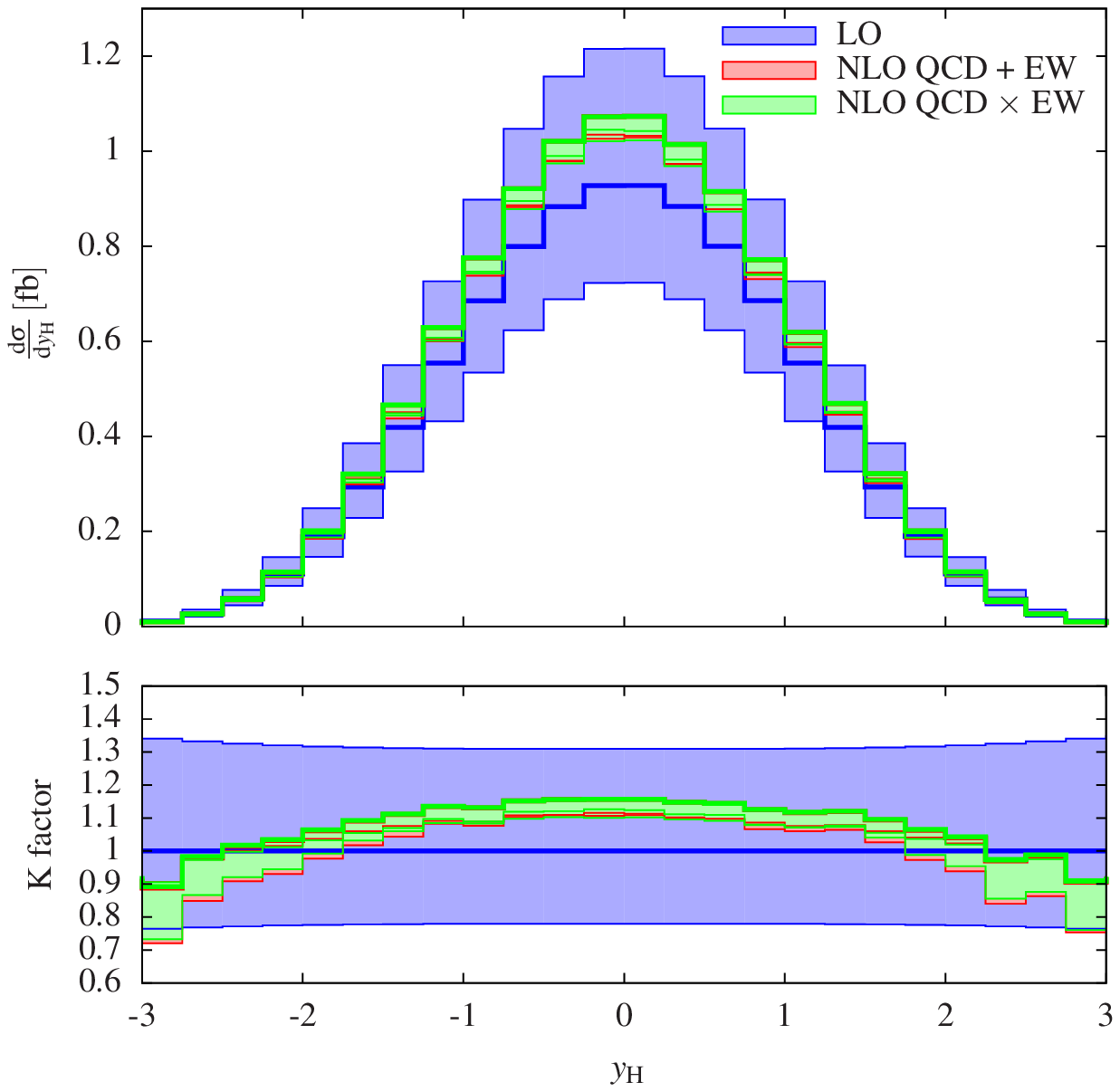}
                \label{plot:rapidity_higgs_qcdew} 
        \end{subfigure}

        \vspace*{-3ex}
        \caption{\label{fig:various_distributions_combined}%
          Combined NLO EW and QCD corrections for various
          distributions at the LHC running at a centre-of-mass energy $\sqrt{s}=13\TeV$: 
          \subref{plot:invariant_mass_truth_top_qcdew}~invariant mass of the reconstructed top quark~(upper left), %
                \subref{plot:invariant_mass_truth_ttxh_qcdew}~invariant mass of the reconstructed $\Pt\bar{\Pt}\PH$ system~(upper right), %
                \subref{plot:cosine_angle_separation_epmu_qcdew}~cosine of the angle between the positron and the muon~(lower left) %
                \subref{plot:rapidity_higgs_qcdew}~rapidity of Higgs boson~(lower right).
          In the upper panels the LO, the NLO QCD$+$EW and NLO QCD$\times$EW distributions are shown.
          The lower panels display the two combined NLO predictions with respect to the LO one as $\mathrm{K} = \sigma_\text{NLO} / \sigma_\text{LO}$.}
\end{figure}

In \reffi{fig:various_distributions_combined} invariant-mass and
angular distributions are presented.  In
\reffi{plot:invariant_mass_truth_top_qcdew}, \change{a radiative tail
  resulting from} non-recombined QCD-partons/photon is perfectly
visible in the invariant-mass distribution of the reconstructed top
quark.  At $167\GeV$, the effect of the EW corrections was found to be
large (about $+15\%$ as can be seen in
\reffi{fig:various_differential_distributions}a).  In the radiative
tail NLO QCD$+$EW and NLO QCD$\times$EW differ by $50\%$.  Where EW
(and QCD) corrections are large, some higher-order contributions to
mixed QCD-EW contributions might still be relevant.

The invariant mass of the reconstructed t$\bar{\rm t}$H system, displayed in \reffi{plot:invariant_mass_truth_ttxh_qcdew}, shows the combined effect of the NLO QCD and EW correction as a function of the partonic centre-of-mass energy.
Both QCD and EW corrections diminish the cross section towards higher energies.
In the distribution of the cosine of the angle between the positron and the muon (\reffi{plot:cosine_angle_separation_epmu_qcdew}), 
both NLO corrections give increasing contributions between $\cos \theta_{e^+ \mu^-} = -1$ and $\cos \theta_{e^+ \mu^-} = 1$ ranging from $0\%$ to $30\%$.
Finally, the distribution in the rapidity of the Higgs boson is shown in \reffi{plot:rapidity_higgs_qcdew}.
This prediction is basically dominated by the QCD effects as the EW corrections do not show any noticeable shape distortion over the whole range for this observable.
This holds true as well for other rapidity distributions.

To conclude, the combined predictions at NLO QCD and EW are, as
expected, mostly dominated by QCD effects.  Nonetheless, in some
phase-space regions EW corrections are non-negligible and should be
taken into account in any precise analysis.  The difference between
the additive and multiplicative combination turns out to be small
apart for some observables where EW corrections are large.
\change{In particular, it is always smaller than the scale uncertainty of the NLO QCD corrections}.

\section{Conclusions}
\label{sec:Conclusions}

The production of the Higgs boson in association with two top quarks will soon be measured at the LHC.
In that respect precise predictions directly comparable with experiments are of prime importance.
This means that decay products of the top quarks should be included, and event selections should be applied to them.
In that way, one obtains realistic predictions for the full process $\Pp\Pp\to\Pe^+\nu_\Pe \mu^-\bar{\nu}_\mu\Pb\bar{\Pb} \PH$.

For this process, the NLO QCD corrections have been recently computed,
but the NLO electroweak (EW) ones were still missing.  We have filled
this gap by computing NLO EW corrections for the full process.  This
constitutes a particularly challenging computation especially for the
virtual corrections.  There, for the first time in a public NLO
computation, one-loop amplitudes featuring up to 9-point functions
appear.  This is rendered possible by the use of the computer codes
\recola and \collier that can provide fast and reliable tree and
one-loop amplitudes particularly suited for Monte Carlo simulations.

Our calculation of EW corrections to the full process
$\Pp\Pp\to\Pe^+\nu_\Pe \mu^-\bar{\nu}_\mu\Pb\bar{\Pb} \PH$ includes
all interference, off-shell, and non-resonant contributions.  The EW
corrections to the fiducial cross section turn out to be small,
\emph{i.e.}~below one per cent.  Nonetheless, in certain phase-space
regions, the EW corrections do become large. They can reach
$-10\%$ for some transverse-momentum distributions, \emph{e.g.}~the
missing-transverse-energy distribution.  The enhanced electroweak
corrections are due to Sudakov logarithms that grow negatively large
when all invariants are large.
\change{Other effects such as radiative tails resulting from real
  photon emission can also be sizeable.} 
This occurs for example for the distribution in the invariant mass of
the reconstructed top quarks where the EW corrections amount to $15\%$.
In the end, the inclusion of NLO EW corrections is mandatory in any
precise analysis of Higgs production in association with two top
quarks.

As the aforementioned computation is particularly challenging, we have
supplemented it with two pole approximations.  For the first
approximation, two resonant top quarks are demanded, while in the
second one two WW gauge bosons are required.  Beyond checking the
full computation, this also allows to investigate the impact of
non-resonant contributions.
Our results for the comparison are consistent with previous observations for the NLO EW corrections to the off-shell
production of a top--antitop quark pair.  
When demanding only two resonant top quarks, errors with respect to the full calculation can reach $4\%$ at LO.
On the other hand, requiring two W bosons seems to constitute a solid
approximation.  In all the distributions that we have computed, this
approximation applied at LO has never been deviating by more than $2\%$ with respect to the full calculation.
At NLO, when using the double-pole approximation only for the subtracted virtual corrections, the differences to the full calculation stay below $1\%$ for two resonant top quarks or two resonant W bosons.
Nonetheless these conclusions only hold for the distributions that we have checked and for the shown phase-space range.
In general one should rely on the full computation in
order to ensure not to miss any off-shell or non-resonant effects.

In order to provide state-of-the-art predictions at NLO we have
combined our NLO EW computation with an already existing NLO QCD
computation.  In this way, both effects are accounted for in a common and
consistent set-up.  Even if QCD effects are dominant, EW corrections
are still important and should be taken into account.  The additive
and multiplicative combinations of the NLO corrections do not show
large differences apart from some particular phase-space regions.
This difference can be taken as an estimate of the missing
higher-order contributions to mixed EW-QCD corrections.

The production of a Higgs boson in association with a pair of top
quarks will probably soon be measured at the LHC.  Thus, realistic and
precise predictions will become very important.  The present work will
help the experimental collaborations to explore further the properties
of the Higgs boson and maybe discover the existence of new physics.

\acknowledgments 
We are grateful to Robert Feger for providing and supporting the code \mocanlo.
The work of A.D. and M.P. was supported by the Bundesministerium f\"ur Bildung und Forschung (BMBF) under contract no. 05H15WWCA1 and the work of J.-N.L. by the Studienstiftung des Deutschen Volkes.
The work of S.U.\ was supported in part by the European Commission through
the ``HiggsTools'' Initial Training Network PITN-GA-2012-316704.

\bibliographystyle{JHEPmod}
\bibliography{ttxh_nlo} 

\end{document}